\documentclass[5p,number,sort&compress,10pt,a4paper]{elsarticle}
\usepackage{amssymb}
\usepackage{mathrsfs,amsmath}
\usepackage{bm}
\usepackage[toc,page]{appendix}
\usepackage{natbib}
\usepackage{graphicx}
\usepackage{color,soul}
\usepackage{float}
\usepackage{todonotes}
\usepackage{makecell}
\usepackage{placeins}
\usepackage{caption}
\usepackage{subcaption}
\usepackage[export]{adjustbox}
\usepackage{algorithm}
\usepackage{algpseudocode}

\newcommand\blfootnote[1]{%
  \begingroup
  \renewcommand\thefootnote{}\footnote{#1}%
  \addtocounter{footnote}{-1}%
  \endgroup
}

\journal{Computational Materials Science}

\begin{document}

\begin{frontmatter}

\title{Self-supervised feature distillation and design of experiments for efficient training of micromechanical deep learning surrogates}

\author[MSD]{Patxi Fernandez-Zelaia\corref{cor1}}\ead{fernandezzep@ornl.gov}
\author[MSD]{Jason Mayeur}\ead{mayeurjr@ornl.gov}
\author[MSD]{Jiahao Cheng}\ead{chengj@ornl.gov}
\author[CSED]{Yousub Lee}\ead{leey@ornl.gov}
\author[SE]{Kevin Knipe}\ead{kevin.knipe@siemens-energy.com}
\author[SE]{Kai Kadau}\ead{kai.kadau@siemens-energy.com}

\address[MSD]{Manufacturing Science Division, Oak Ridge National Laboratory, Oak Ridge, TN, United States}
\address[CSED]{Computational Sciences \& Engineering Division, Oak Ridge National Laboratory, Oak Ridge, TN, United States}
\address[SE]{Siemens Energy, Orlando, FL, United States}

\cortext[cor1]{Corresponding author}

\begin{abstract}

Machine learning surrogate emulators are needed in engineering design and optimization tasks to rapidly emulate computationally expensive physics-based models. In micromechanics problems the local full-field response variables are desired at microstructural length scales. While there has been a great deal of work on establishing architectures for these tasks there has been relatively little work on establishing microstructural experimental design strategies.  This work demonstrates that intelligent selection of microstructural volume elements for subsequent physics simulations enables the establishment of more accurate surrogate models. There exist two key challenges towards establishing a suitable framework: (1) microstructural feature quantification and (2) establishment of a criteria which encourages construction of a diverse training data set. Three feature extraction strategies are used as well as three design criteria. A novel contrastive feature extraction approach is established for automated self-supervised extraction of microstructural summary statistics. Results indicate that for the problem considered up to a 8\% improvement in surrogate performance may be achieved using the proposed design and training strategy. Trends indicate this approach may be even more beneficial when scaled towards larger problems. These results demonstrate that the selection of an efficient experimental design is an important consideration when establishing machine learning based surrogate models.

\end{abstract}

\begin{keyword}
machine learning \sep experimental design  \sep ICME \sep surrogate modeling  \sep micromechanics \end{keyword}

\end{frontmatter}

\blfootnote{\textbf{Notice of Copyright}. This manuscript has been authored by UT-Battelle, LLC under Contract No. DE-AC05-00OR22725 with the U.S. Department of Energy. The United States Government retains and the publisher, by accepting the article for publication, acknowledges that the United States Government retains a non-exclusive, paid-up, irrevocable, world-wide license to publish or reproduce the published form of this manuscript, or allow others to do so, for United States Government purposes. The Department of Energy will provide public access to these results of federally sponsored research in accordance with the DOE Public Access Plan (http://energy.gov/downloads/doe-public-access-plan).}

\section{Introduction}
\label{section:intro}

Numerical physics-based models are essential tools needed to study many complex physical systems. However, due to the computational burden associated with discretizing and solving the governing laws, using these models to optimize and design engineering systems is difficult. Driven in part by the materials genome initiative integrated computational materials engineering (ICME) tools have become ubiquitous in various fields \cite{de2014materials}. While classically focused on physics-based simulations and statistical approaches the ICME paradigm has benefited greatly from recent advances in machine learning (ML) and the development of open source frameworks. For instance, original methods for binary phase image generation relied on pixel-wise simulated annealing approaches \cite{torquato2002random}; presently ML-based generative diffusion models can produce much more realistic and complex structures \cite{fernandez2024digital,buzzy2024statistically,robertson2023local,lee2023microstructure}. Google and Microsoft have even recently released materials focused generative models indicating their interest in ICME approaches \cite{zeni2023mattergen,yang2023scalable}.

\begin{figure*}[]
\begin{center}
  \includegraphics[width=1.0\linewidth]{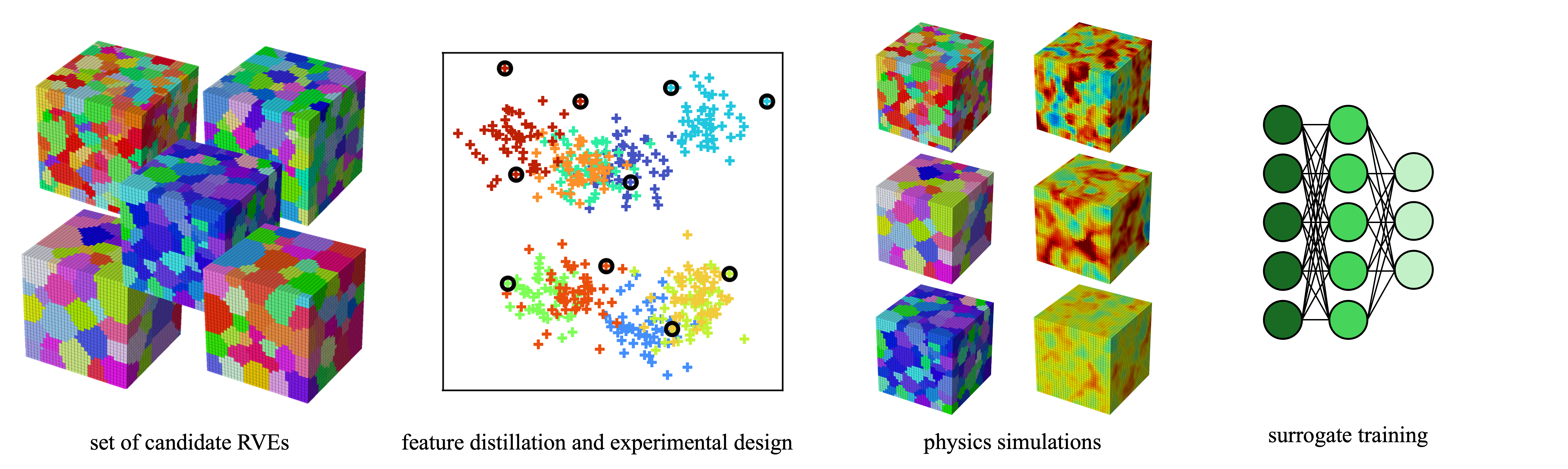}
  \caption{Overall approach is to identify most unique and informative MVEs for subsequent physics evaluation and surrogate training. Hypothesis is that more efficient training may be performed if MVEs are chosen using an appropriate design criteria.}
  \label{fig:overallApproach}
\end{center}
\end{figure*}

In the statistics community classic surrogate reduced order models have been typically used to emulate parameterized numerical codes with scalar valued outputs \cite{sacks1989designs,kennedy2001bayesian,gramacy2020surrogates}. Gaussian processes (GPs) are favored in the field of computer experiments because, for deterministic simulations, GPs can be formulated to behave as \textit{interpolators}; the function will pass exactly through the training data without over fitting. Modern GP implementations have been formulated which can capture more complex input and output structures \cite{mak2018efficient,chen2021function}. Additionally there are composite approaches combining GPs with moderm ML approaches \cite{damianou2013deep,dai2015variational}. Fundamentally GPs can be thought of as making a prediction at an input $\bm{x}$ by performing a weighted average of ``near-by'' training examples. This measure of closeness requires calculation of a distance metric (and learning of spatial correlation functions). For parametric inputs this is a rather simple Euclidean distance for low-dimensional data. Parametric definition of a complex system is made possible via careful use of relevant summary statistics \cite{tran2021solving}. However, for many engineering and science problems parametric description of the problem may not be feasible. For instance, emulating finite element (FE) models with discretized complex geometries, which cannot be easily captured via GPs, can however be emulated using ML approaches \cite{liang2018deep,he2023novel}. Similarly, in many mechanics problems the input to a numerical code may not only be fundamental scalar properties (stiffness, strength) but input \textit{structural} representations. For instance, shown in Fig. \ref{fig:overallApproach}, microstructural volume elements (MVEs) contain $32^3$ voxels with the local state in each being represented by three Euler angles. This corresponds to $32,768\cdot 3$ total ``features'' which need to be considered in computing the necessary GP distance measure. Furthermore, in certain settings the full-field (voxel-wise) response is needed e.g. at least $32,768$ total output values. Hence, for these complex structural problems, direct application of classic GPs is not well suited. Instead, convolutional neural networks (CNNs) have been shown to be well suited towards emulating these micromechanical localization problems \cite{montes2022convolutional,yabansu2014calibrated,khorrami2023artificial,wang2021stressnet,pokharel2021physics,pandey2021machine}. This is because the convolution operation is well suited for processing spatial structure and CNN architectures may be easily designed for multiple output predictions. Furthermore there is a clear link to Green's functions in continuum theory which use analytical convolution operations to predict the response of heterogenous mediums \cite{yabansu2014calibrated}. This theoretical basis provides good justification for using data-driven CNN networks.

Development of effective statistical or ML surrogate models not only requires a suitable model form but also careful construction of an appropriate \textit{experimental design}. In physical experiments designs are often focused on eliminating confounding or biasing factors, mitigating against the effects of experimental noise, and extracting trends e.g. estimating the sensitivity of the response variable to input factors \cite{wu2011experiments}. In the statistics community designs for surrogate model development favor \textit{space-filling} designs \cite{joseph2016space}. Designs here refer to the collection of training examples $\left\{\bm{x}_1,\ldots,\bm{x}_N\right\}$ used to evaluate the physics model and train the surrogate. Consider that points close to one another are assumed to have similar responses. In fact, for deterministic simulations, in the limit as two points overlap the responses are identical. Hence, space-filling designs are constructed by optimizing a specified criteria which encourages the establishment of a desirable distribution of points in space. The criteria always requires definition of a distance metric as it is necessary to avoid clustering. There exists a number of different kinds of space-filling design criteria each consisting of trade offs between design quality and optimization complexity \cite{joseph2016space}.

Development of designs for the micromechanics problem, shown in Fig. \ref{fig:overallApproach}, is extremely challenging due to the need to define a distance metric for pair-wise comparison of structures. In certain materials problems this is made possible via the use of domain-science knowledge to define statistical descriptors.  For instance in composite structures features may include  phase volume fraction, particle size, particle shape, etc. \cite{bessa2017framework}. For polycrystalline systems diversity of crystallographic texture is critically important \cite{paulson2017reduced}. A recent work has established a homogenization framework which predicts aggregate constitutive properties using single-crystalline responses \cite{he2024material}. Prior localization works have engineered MVEs to effectively train surrogates \cite{yabansu2014calibrated}; single-voxel grains for fine scale responses, multi-voxel grains for larger scales, and single crystals with embedded single-voxel speckles to capture ``delta'' localization phenomena. There are a number of works that have demonstrated that diversity of training data in atomistic problems is of paramount importance as well \cite{montes2022training,barneschi2024molecular}. In one work an entropy-based criteria was used to increase the volume of the training data to mitigate against extrapolation \cite{montes2022training}. Critically, however, there are no existing works that quantitatively demonstrate the importance of the experimental design in training full-field micromechanical ML surrogate models.

This work hypothesizes that intelligent selection of MVEs for subsequent physics simulations will enable the establishment of more accurate ML surrogate models. The overall strategy, shown in Fig. \ref{fig:overallApproach}, is similar to that of traditional computer experiments; distribute settings over a diverse space, run simulations, and train a surrogate model. The main challenge, however, is that it is not immediately straight forward how to quantify the MVE inputs. Hence, three approaches are used to distill the MVEs to a lower dimensional representation: (1) a variational autoencoder (VAE) latent description (2) self-supervised derived microstructural statistics and (3) domain science inspired microstructural statistics. Furthermore, due to differences in information contained in each descriptor three space-filling design criteria are parametrically tested to evaluate their performance. When compared against a random selection of MVE training examples results indicate that up to a 8\% boost in surrogate performance, shown to be statistically significant, may be achieved via the proposed training strategy. We observed that this benefit increases with increasing data set size, and hence, this approach may be even more beneficial when scaled towards larger problems. Interestingly, it is observed that while the mean performance of the surrogate is only marginally sensitive to the microstructure, there are significantly more poorly performing outliers for textured large grain MVEs. We suspect that this may simply be due to smaller grain size instances being more representative and, hence, containing more information. These results demonstrate that experimental design for micromechanical problems is imperative. In addition, our self-supervised approach for distilling microstructure statistics may be suitable for other tasks where image similarity metrics are important.

\section{Methods}
\label{section:methods}

It is hypothesized here that careful selection of MVEs, for subsequent physics simulation, will enable training of a more accurate micromechanical deep learning surrogate when compared against random selection. In the active learning context it is assumed that a trained ML model is already available and activations from neurons may be used as features when constructing the sequential design e.g. identifying subsequent examples for labeling, physics simulation, etc. \cite{sener2017active}. This approach is feasible as targeting novel activations when constructing the sequential design will indeed ensure that the model will ``see'' novel examples during subsequent training. However, there are a few key challenges which arise which may be problematic for the materials problem posed here. In ML the active learning sequential design task is often referred to as core-set and the design is optimized using variants of the maximin distance criteria. In the referenced work the VGG-16 architecture is used on CIFAR-10 and SVHN data sets (60,000 and 600,000 images in each, respectively). These authors note that activations from the final fully connected layer are used which is 1,000-dimensional. At such high dimensions the search space becomes extremely large and constructing a design is challenging. For natural images this may be less of a concern as data sets are rather large (60,000 and 600,000 in this case, some image data sets 1M+). In our context, with only a total of 6,825 MVEs, searching a 1000-dimensional space would be extremely challenging. Furthermore, the challenge addressed here is focused on the initial design of experiments and, hence, there is no ML available from which features may be extracted.

Despite the initial design challenge being fundamentally different than the active learning problem the two approaches share similar strategies; identify salient features, use a design criteria to identify a diverse design, perform physics simulations (or perform labeling in the natural image context), and train the ML model. In a CNN activations can be related to \textit{localized} features. For instance, a candidate example which exhibits distinctly different activations than examples already in the training data set may contain unique edges, corners, shapes, textures, etc.. Alternatively, unique activations may correspond to similar features but at different spatial locations in the input. In the micromechanical surrogate model this seems appealing; a data set containing various triple junctions of different orientation relationships will be desirable towards training a generalizable model. Conversely, the same triple joint embedded in two volumes of differing overall crystallographic texture may behave differently. Therefore, both statistical features which describe aggregate qualities of the microstructure, and localized features which describe spatial arrangement of structure, are likely to provide value.

Three key design criteria will be used for assessing the diversity of training ensembles. In the context of computer experiments \textit{space-filling designs} are commonly used to ensure good ``spread'' in the $d$-dimensional model input space \cite{joseph2016space}. The challenge in identifying a design is in minimizing the corresponding criteria; consider a design of $N$ $d$-dimensional points contains $N\cdot d$ total values which must be identified. Furthermore, the curse of dimensionality drives search spaces to be ever larger with increasing $d$. Three key space-filling design criteria will be considered in this work; a greedy maxi-min design \cite{joseph2016space}, a greedy maximum projection (maxPro) design \cite{joseph2015maximum}, and a data twinning approach \cite{vakayil2022data}.

\subsection{Physics simulation}
\label{section:physics}

The local stress response for input MVEs was simulated using open source software \textit{PRISMS-plasticity} \cite{yaghoobi2019prisms}. In this exploratory work the material was assumed to behave elastically so that a sufficiently large data set could be quickly generated for testing the stated hypothesis. While this work is specifically focused on the simplified elastic constitutive response the main contribution here is demonstrating the importance of the initial experimental design; this should extend to any mechanical constitutive model. Elastic constants corresponding to a Ni-based superalloy were utilized with $C_{11}=199\,GPa$,  $C_{12}=128\,GPa$, and $C_{44}=99\,GPa$ \cite{fernandez2022creep}. MVEs were loaded in uniaxial tension with a $50\,MPa$ traction applied on the top $z=1$ plane and periodic boundary conditions elsewhere. A $32 \times 32 \times 32$ element mesh was used with linear interpolation shape functions. Simulations were run over \textit{all} candidate MVEs to assemble a large data set. Experimental designs were tested by sub-sampling from this data set prior to training the surrogate models.

\subsection{Microstructural features}
\label{section:feature_extraction}

The microstructural MVEs used in this work were generated using open source software \textit{Neper} \cite{quey2022neper}. Each example is $32^3$ with grain sizes varying from 4 to 16 voxels. A total of 6,825 examples were generated. 1,200 of these (12 grain sizes, 100 total random seeds) were generated with uniformly random crystallographic texture. The remaining MVEs were generated with a randomly selected $(hkl)$ fiber texture in a randomly selected $<uvw>$ direction. Furthermore, to avoid the presence of solely sharp textures, orientations were subsequently diffused by randomly applying rotations with Euler angles $\sim Unif(-10^{\circ},10^{\circ})$. This randomization ensures that both sharp and diffuse textures are present in the generated data set.

Three procedures for quantifying microstructural features will be considered in this work: latent space features from a VAE, features from a novel self-supervised network, and classical microstructural descriptors. For the classical microstructural descriptors the grain size, prescribed during generation, is combined with volume-averaged crystallographic information. The latter is captured using generalized spherical harmonics (GSH) \cite{bunge2013texture}. The orientation distribution function at each spatial location, $\bm{x}$, can be described by a basis expansion

\begin{equation}
\begin{aligned}
f_{\bm{x}}\left(\bm{g}\right) = \sum_{\mu,n,l} F_{l\bm{x}}^{\mu n} \dot{\dot{T}}_l^{\mu n}\left(\bm{g}\right),
\label{eqn:GSH}
\end{aligned}
\end{equation}

\noindent
where $\bm{g}$ are the Euler angles, $\mu,n,l$ are indices for multiple sums, and $F_{l\bm{x}}^{\mu n}$ are the GSH coefficients at $\bm{x}$. An expansion consisting of nine total terms was used which has been shown to be sufficient for similar quantitative tasks for FCC materials \cite{yabansu2014calibrated}. $\dot{\dot{T}}_l^{\mu n}$ are the corresponding GSH basis functions. Both basis weights and functions are complex valued. The orientation distribution function over a volume can be obtained by simply computing the mean over all spatial locations of the individual basis coefficients. As this representation is complex in nature, real and imaginary components are taken and concatenated together with grain size information to define a ``classic'' descriptor $\bm{z}\in \mathcal{R}^{18}$.

\begin{figure}[]
\begin{center}
  \includegraphics[width=1.0\linewidth]{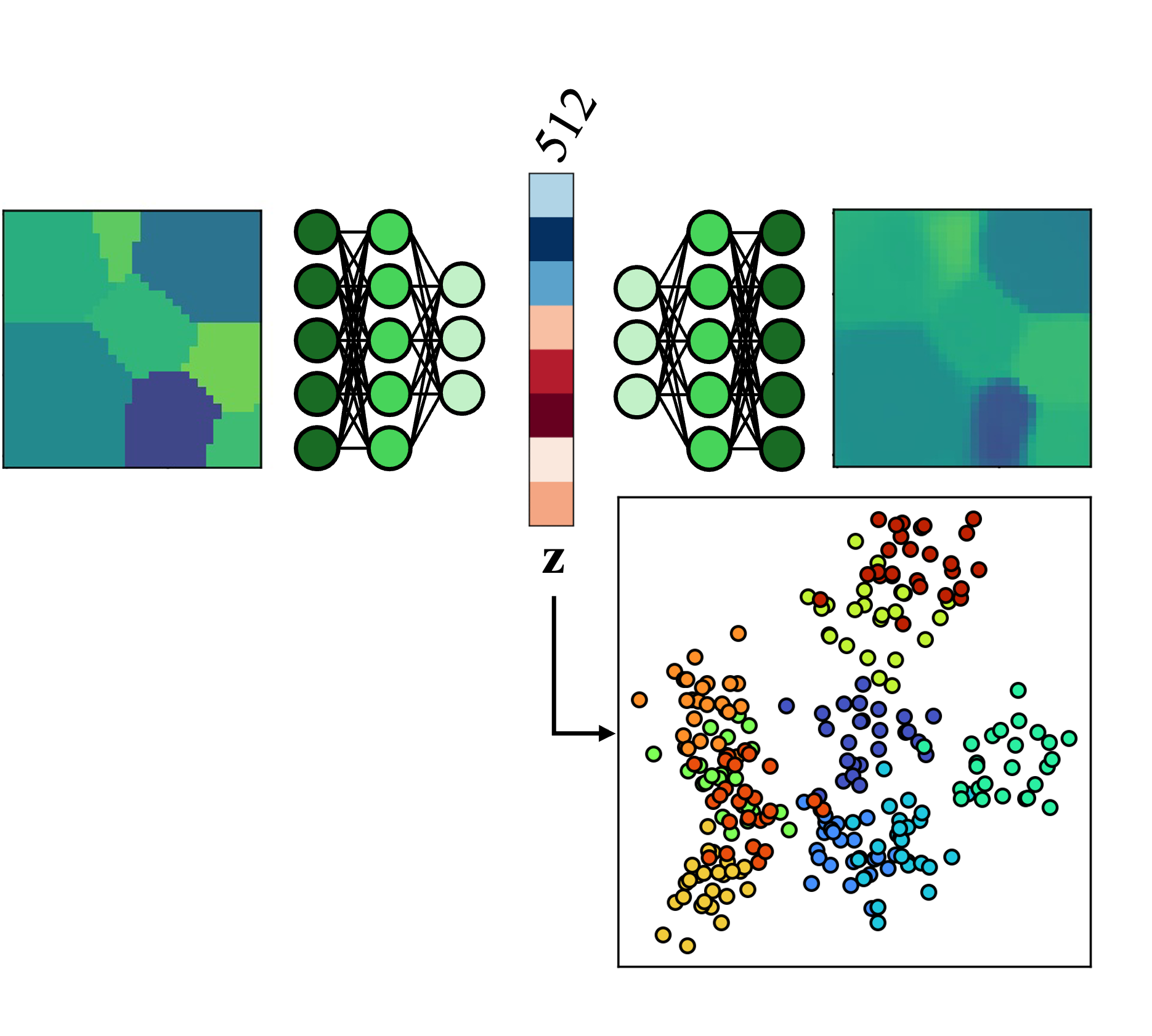}
  \caption{VAE schematic for extracting localized MVE features.}
  \label{fig:vaeSchematic}
\end{center}
\end{figure}

VAEs construct feature vectors via a non-linear dimensionality reduction mapping \cite{goodfellow2016deep}. Shown in Fig \ref{fig:vaeSchematic} is a schematic of inputs, outputs, and the latent representation. The network consists of an encoder, which maps input $\bm{x}$ to a latent $\bm{z}$, and a decoder which maps back to the original modality $\hat{\bm{x}}$. The network is trained in an unsupervised fashion to minimize the discrepancy between input and outputs e.g. minimize the reconstruction loss. While similar to autoencoders VAEs also include additional regularizing constraints on the latent space. First stochasticity is introduced in the model by having the encoder predict both a mean latent representation and a corresponding variance measure. Next the ``reparameterization trick'' is used to sample from this latent distribution; white noise is scaled by the networks variance prediction and added to the predicted mean. This perturbed latent vector is mapped back to $\hat{\bm{x}}$ and the mean and variance parameters contribute to an additional Kullback–Leibler (KL) divergence loss term that encourages the latent space to be a standard normal distribution. This effort is needed to ensure that the latent space is continuous. Consider that small latent perturbations in an AE may produce a nonsensical output e.g. values may not be interpolated in the latent space. Since the VAE uses a distribution to represent the latent space it intrinsically includes small perturbations in the training procedure e.g. the $\bm{x}\rightarrow \bm{z}$ mapping never produces the same $\bm{z}$ but nonetheless close $\bm{z}$'s. This ensures that the latent is continuous and may indeed be interpolated. This is important since this work seeks to use distance measures for experimental design and, hence, it is important that the latent space be continuous in some fashion. Furthermore, a variation of the VAE, $\beta$-VAE, was used to balance the KL and reconstruction loss terms \cite{higgins2017beta}. For extracting features a fairly large latent space was necessary to produce reasonable reconstructions e.g. $\bm{x}\in \mathcal{R}^{512}$.  Note that the original dimensionality of each MVE is $32^3 \cdot 3 = 98,304$ and so this still represents a compression ratio of $\sim 200$. The high dimensionality of the latent space may be valuable in terms of quantifying localized spatial features, however, this may present problems when constructing designs (high dimensional optimization problem). 

\begin{figure*}[]
\begin{center}
  \includegraphics[width=1.0\linewidth]{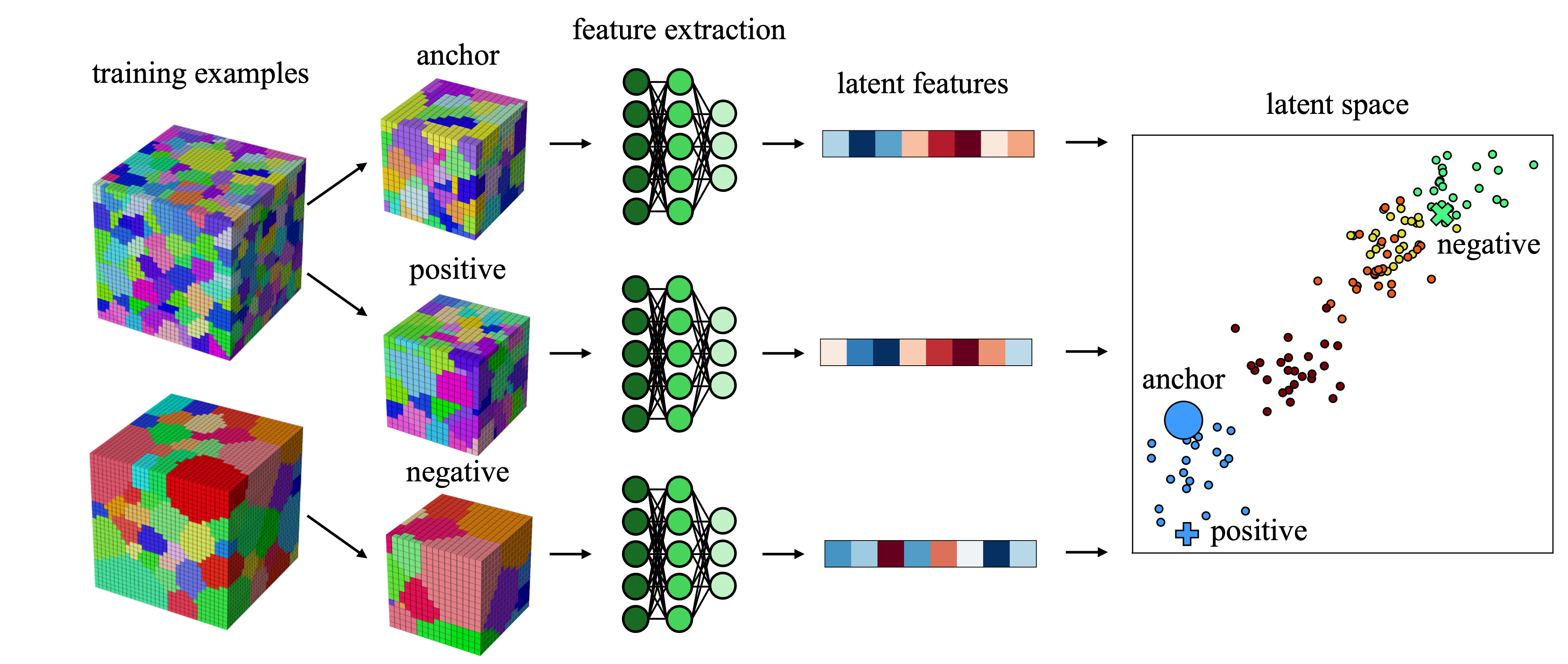}
  \caption{Self-supervised approach for 3D MVE feature extraction and network training. Subsampling of MVEs enables self-supervised learning and, critically, encourages the learning of statistical descriptors.}
  \label{fig:selfSupervisedSchematic}
\end{center}
\end{figure*}

Finally a self-supervised alternative is proposed for automatically extracting relevant microstructural statistics from MVEs. Materials at most length scales are inherently stochastic. Consider, for instance, that two micrographs from the same material will have different appearance but statistically may be identical. A VAE may not capture this subtlety as it is trained to optimize a reconstruction and, hence, explicitly captures localized features. As an example consider that two nearly identical micrographs, offset by a small horizontal translation, will produce two different $\bm{z}$'s when passed through the encoder network. Hence, there is a need to establish an automated way to infer summary statistics which can enable pair-wise similarity measurement across stochastic MVEs. Interestingly, with a few exceptions, there are very few works in the materials community focused on using ML for these kinds of tasks \cite{hashemi2024toward}. Perhaps this is because most ML vision models are focused on natural images, where localization is important, and, hence, there are fewer models suitable for the stochastic materials problem. Perhaps the closest related task occurs in the field of texture analysis (not crystallographic texture but image texture) \cite{larroza2016texture}. Interestingly there is a similar implementation, focused on the extraction of statistical features from a CNN, for remote sensing imagery \cite{liu2019statistical}. In biological applications there are also extensive applications of ``twin'' networks for measuring similarity between micrographs \cite{presberger2024correlation}. In summary, there is a critical need for specialized ML methods applicable and tailored for extraction of statistical features from stochastic materials data.

Contrastive learning seeks to establish ML models via training on pair-wise similarity measures \cite{chopra2005learning,goldberger2004neighbourhood,bromley1993signature,balestriero2022contrastive,hadsell2006dimensionality}. This is valuable for the experimental design problem as all design criteria rely on distance metrics. A schematic of the self-supervised contrastive learning procedure used in this work is shown in Fig \ref{fig:selfSupervisedSchematic}. During training a $32^3$ MVE is sub-sampled to produce two $16^3$ MVEs; an anchor ($\bm{x}_a$) and a positive ($\bm{x}_p$) example. Then a negative ($\bm{x}_n$) $16^3$ MVE example is randomly cropped from another random training example. Since the anchor and positive examples come from the same parent MVE their statistics should be similar. This of course assumes stationary behavior and that the $16^3$ MVE is representative. Likewise the statistics of the anchor and negative example will, on average, be different. Statistics are generated via training of a single NN which maps each example into latent space vector. In the latent space a contrastive loss is defined which encourages the network to place the anchor and positive examples close together and the negative example far apart. The training loss can be defined as \cite{hadsell2006dimensionality},

\begin{equation}
\begin{aligned}
\mathcal{L} = \max (0,d(\bm{x}_a,\bm{x}_p)-d(\bm{x}_a,\bm{x}_n)+1/2),
\label{eqn:contrastiveLoss}
\end{aligned}
\end{equation}

\noindent
where $d(\cdot,\cdot)$ is the Euclidean distance. The constant $1/2$ is referred to as the margin and controls the spacing between points and, hence, the overall scale of the latent space. This loss drives the distance between anchor and positive examples to be as small as possible and, conversely, the distance between the anchor and negative example to be large.

\begin{figure}[]
\begin{center}
  \includegraphics[width=1.0\linewidth]{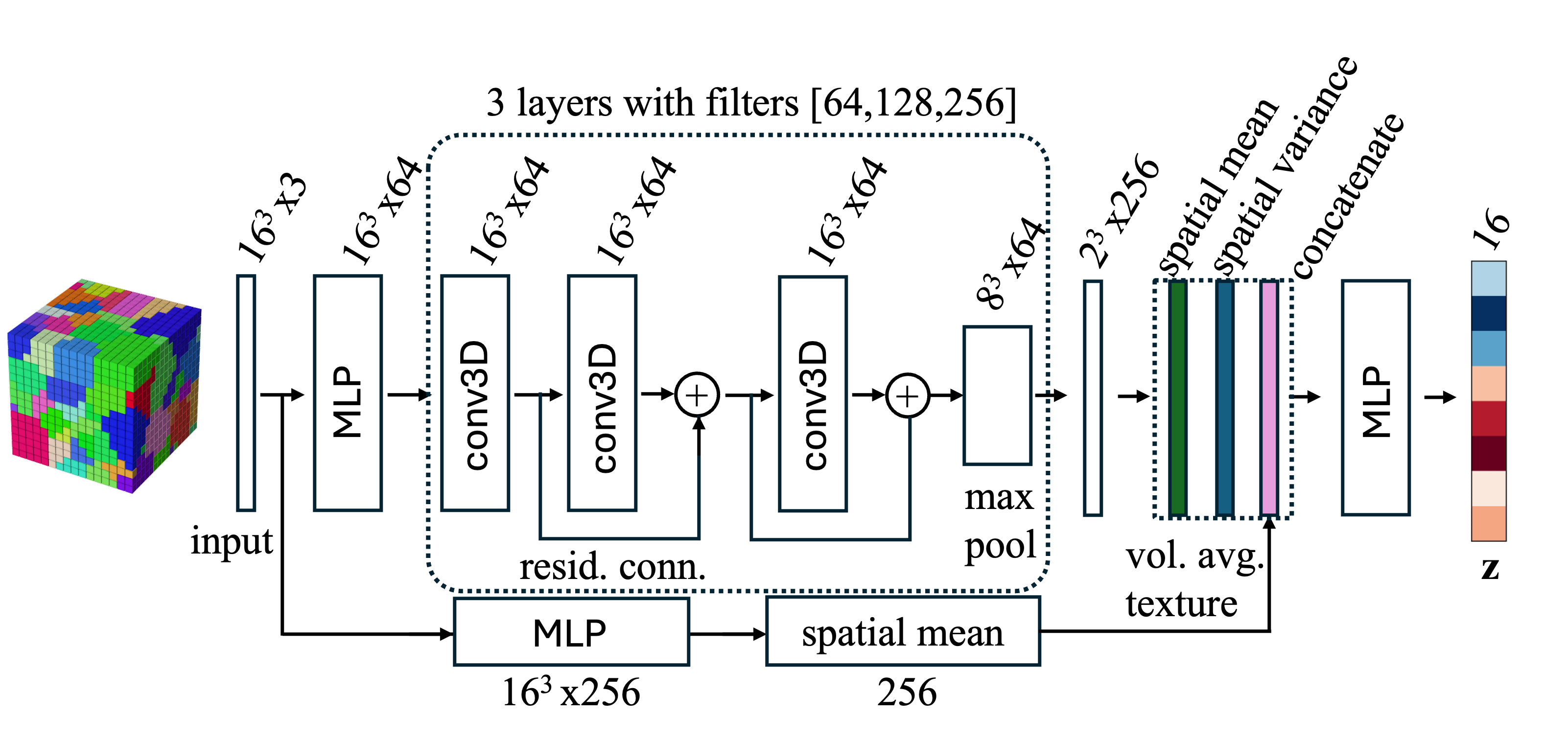}
  \caption{Novelty of the feature extraction procedure is that it intrinsically operates on image \textit{statistics} via construction of the network. Mean and variance of spatial-orientation feature maps are combined with volume averaged orientation features prior to passing through the final MLP. This encourages the network to separately construct orientation and spatial-orientation statistics prior to mixing in the final MLP.}
  \label{fig:selfSupervisedNetwork}
\end{center}
\end{figure}

The specific network architecture used for extracting microstructural statistics is shown in Fig. \ref{fig:selfSupervisedNetwork}. Using domain knowledge, that both local spatial information and volume averaged texture information are important, the network has been tailored to capture these two considerations explicitly. In parallel the network (1) performs a series of CNN operations to construct spatial feature maps and (2) uses an MLP to map the input Euler angle representation to ``texture features''. Spatial averaging is performed on the texture features to produce a global crystallographic feature vector. Similarly, feature map statistics are produced by computing the mean and variance over all spatial indices. These three vectors (mean of spatial-orientation feature maps, variance of spatial-orientation feature maps, and volume averaged orientation features) are concatenated and passed through a MLP to mix the information prior to outputting a $\bm{z}\in \mathcal{R}^{16}$ latent representation. Originally we did not consider the spatial mean/variance pooling operations but found that model performance improved doing so. Furthermore, we originally did not include a separate texture-only information stream but found that this also improved performance. It is suspected that this is because the network is not tasked with ``doing everything'' at once and introduction of this domain knowledge makes training easier.

\subsection{Design of experiments}
\label{section:DoE}

Three key space-filling designs will be considered in this work: maximin distance design \cite{joseph2016space}, maximum projection design \cite{joseph2015maximum}, and a data twinning design \cite{vakayil2022data}. The former two are solely space-filling designs seeking to spread points uniformly in space with some subtle differences. The latter criteria is unique in that it seeks to balance a space-filling objective while also optimizing an distributional objective. The distributional objective seeks to identify design points from a candidate data set which collectively emulate the probability distribution representative of the original data set e.g. the empirical distribution.

\begin{figure}[]
\begin{center}
  \includegraphics[width=1.0\linewidth]{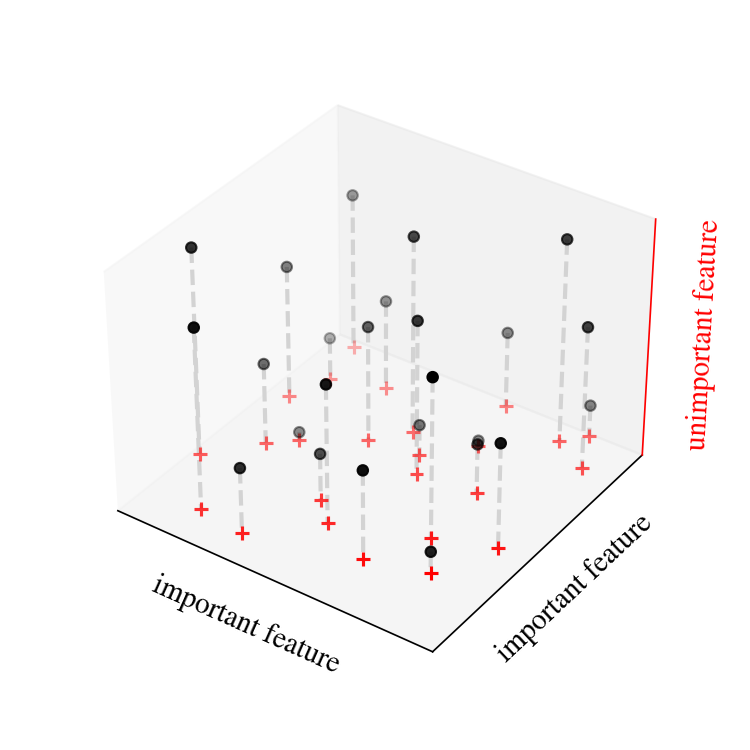}
  \caption{Maximum projection design criteria ensures good spreading in all possible subspace projections. This ensures that even when unknown unimportant features are present the resulting design still exhibits desirable space-filling properties in the effective lower dimensional space.}
  \label{fig:maxProExample}
\end{center}
\end{figure}

The FE simulations performed on MVEs are deterministic in nature therefore it is assumed that similar MVEs will produce similar responses. 
The idea of space-filling designs is to therefore spread points apart in the input space as efficiently as possible. Close points are undesirable as they may produce similar simulation results therefore wasting computational resources. The maximin distance design is tasked with identifying a design which maximizes the minimum pair-wise distance across all points \cite{johnson1990minimax,joseph2016space}. Specifically,

\begin{equation}
\begin{aligned}
\max_{\mathcal{D}} \min_{i,j} d(\bm{x}_i,\bm{x}_j),
\label{eqn:maximin}
\end{aligned}
\end{equation}

\noindent
where $\mathcal{D}$ is the constructed design (collection of $\bm{x}$'s) and $d$ is the Euclidean distance. Here we adopt a \textit{conditional} maximin (cMm) approach where a greedy algorithm is used to select points from a candidate set one at a time. First an initial random point is selected followed by addition of sequential points which step-by-step minimize the objective function. This is sub-optimal but is relatively simple to implement. Note that due to its construction this criteria tends to push points towards corners in the design space.

One possible deficiency of the maximin design is that it does not account for \textit{projections} of the design onto subspaces e.g. x-y-z onto the x-y plane. In experimental design there is a concept of \textit{effect sparsity} which assumes that some features or settings may be unimportant \cite{wu2011experiments}. For instance, consider a maximin design in three dimensions where, unknown to the scientist or engineer, the third variable is unimportant. The three dimensional problem collapses to a two dimensional problem and there is no guarantee that design points are space-filling in this projection e.g. there may be overlap. This deficiency was addressed the maximum projection (maxPro) design \cite{joseph2015maximum},

\begin{equation}
\begin{aligned}
\min_{\mathcal{D}} \sum_i^{n-1} \sum_{j=i+1}^n \frac{1}{\prod_{l=1}^p |d^k(\bm{x}_{il},\bm{x}_{jl})|^2},
\label{eqn:maxPro}
\end{aligned}
\end{equation}

\noindent
where $p$ is the dimensionality of the variable. This objective function includes a product across all dimension-wise distances which penalizes overlap in all possible projections of the data. An example three dimensional design and projection schematic is shown in Fig. \ref{fig:maxProExample}.

\begin{figure}[]
\begin{center}
  \includegraphics[width=1.0\linewidth]{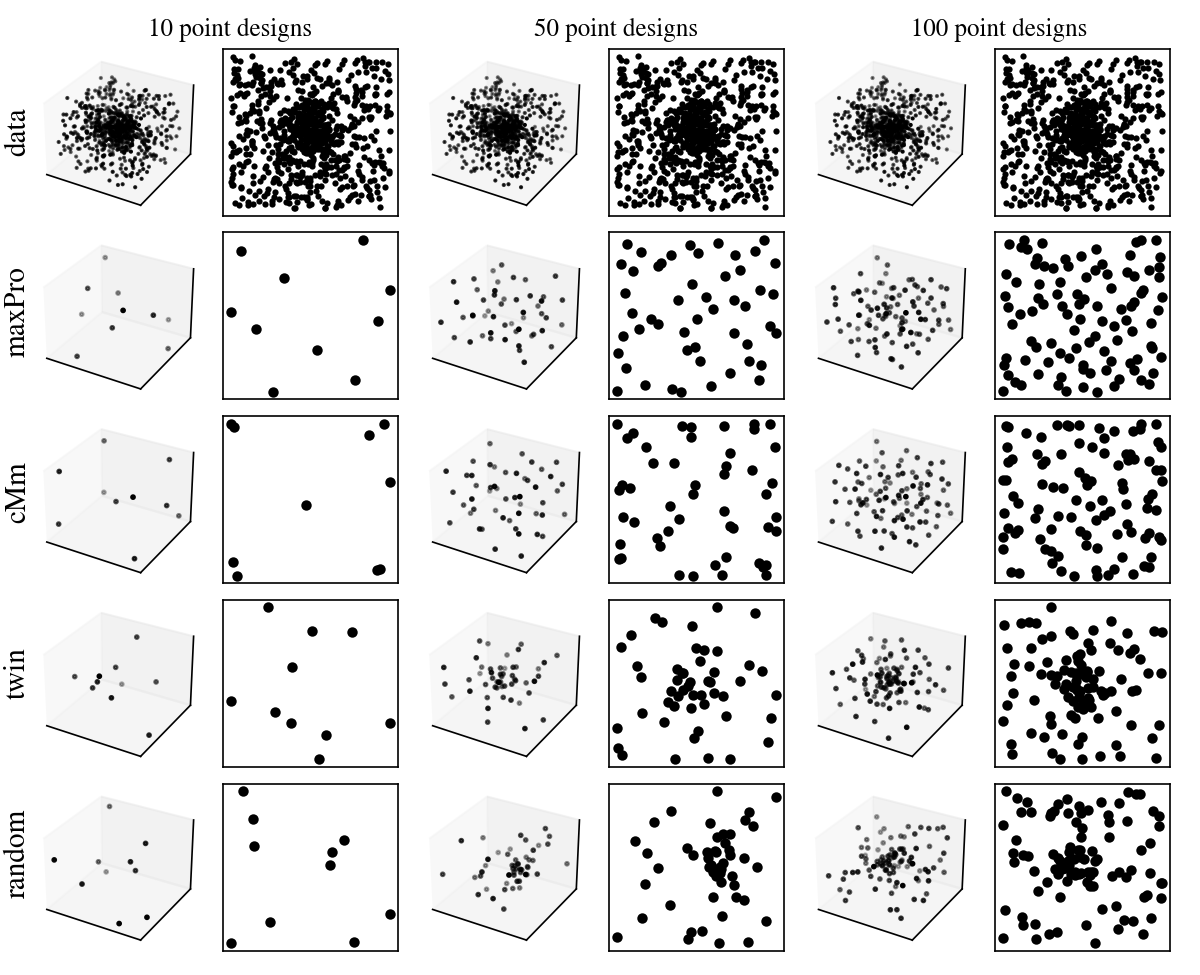}
  \caption{Example designs created from a three dimensional candidate data set consisting of 1000 points from $Unif(-5,5)$ and 500 points from $\mathcal{N}(0,1)$. The three dimensional design and one two-dimensional projection is shown.}
  \label{fig:designExamples}
\end{center}
\end{figure}

Finally the third design criteria considered is a \textit{data twinning} approach \cite{vakayil2022data,mak2018support}. This approach seeks to subsample the candidate data set and identify points which ensures both good coverage of the input space and emulation of the parent data set density. The optimization problem is beyond the scope of this work and readers are referred to the original works. In the context of the present problem data twinning may be useful if (1) the data set is non-uniform e.g. there are more examples of one crystallographic texture than another and (2) if the neural network can be tailored to perform better in these possible high density regions. The latter point is important because the surrogates will be evaluated against a validation dataset; if the validation data set is biased towards certain microstructures then it is beneficial to tailor the training data set and model to be accurate on those structures. This latter point is highly problem specific and even philosophical. The end user must decide if they prefer uniform coverage or preference for more likely structures. Furthermore, it is not clear if the second point is true for the micromechanical surrogate model considered here. GPs behave as interpolators and, hence, having a higher density of points in high density regions will selectively improve model performance near this space. CNNs may not necessarily behave this way. Nonetheless this design criteria will be considered for evaluation.

A three dimensional example illustrating these design strategies is shown in Fig. \ref{fig:designExamples}. Three data set sizes are considered to capture each design criteria's sensitivity to size. Three dimensional scatter plots and a two dimensional x-y projection are shown to demonstrate the effects of the effect sparsity principle. Visually it is clear that random does a rather poor job generally. Large samples are needed to eventually obtain a uniform distribution in both three dimensions and the subspace projection. While cMm does efficiently fill the 3D space nicely there is significant overlap in points when projected to the x-y plane. This is most pronounced in the 10-point design where the criteria pushes nearly all points into corners. Consider that this problem compounds even further in higher dimensions; there are $2^d$ corners in a $d$-dimensional space. The maxPro designs perform uniformly well in all cases with particularly good performance in the x-y projection. Finally, twinning performs as expected by balancing spreading of points along with capturing the distribution of the original candidate data set.

\subsection{Surrogate model architecture}
\label{section:surrogate}

A U-net architecture was used to emulate the micromechanics FE model and predict all six components of the stress tensor. We found that predicting all six components was more effective than simply predicting the von Mises stress. This is possibly be due to correlations between the outputs which provides the network with additional information and constraint during training. Inputs to the model are three-dimensional microstructural representations represented by a $[32,32,32,3]$ array with orientation information encoded via Euler angles. The U-net architecture consists of three resolution depths ($32^3,16^3,8^3$) with each resolution using ($64,128,256$) filters. At each resolution, during both down-sampling and up-sampling, there are four layers of three-dimensional CNNs producing feature maps. Residual connections are used throughout to enable gradient flow \cite{he2016deep}. Batch normalization and dropout layers ($p=0.05$) are also used throughout. A MLP with dimensions $[256,64,16,6]$ is used to map feature channels at the end of the U-net to the six output stress components. Finally, a $L_2$ penalty weight of 0.0001 was used on all weights and biases throughout the model. We found this to be imperative to avoid over-fitting especially for small data set sizes.

The model was implemented in \textit{Tensorflow} and trained using an Adam optimizer with default parameters, a learning rate of $10^{-3}$, and batch size of 64 \cite{abadi2016tensorflow}. Training was performed on 80GB Nvidia A100 GPUs for a total of 200 epochs.

\section{Results}

\subsection{Microstructural representation}
\label{section:microstructure_rep}

\begin{figure}[]
\begin{center}
  \includegraphics[width=1.0\linewidth]{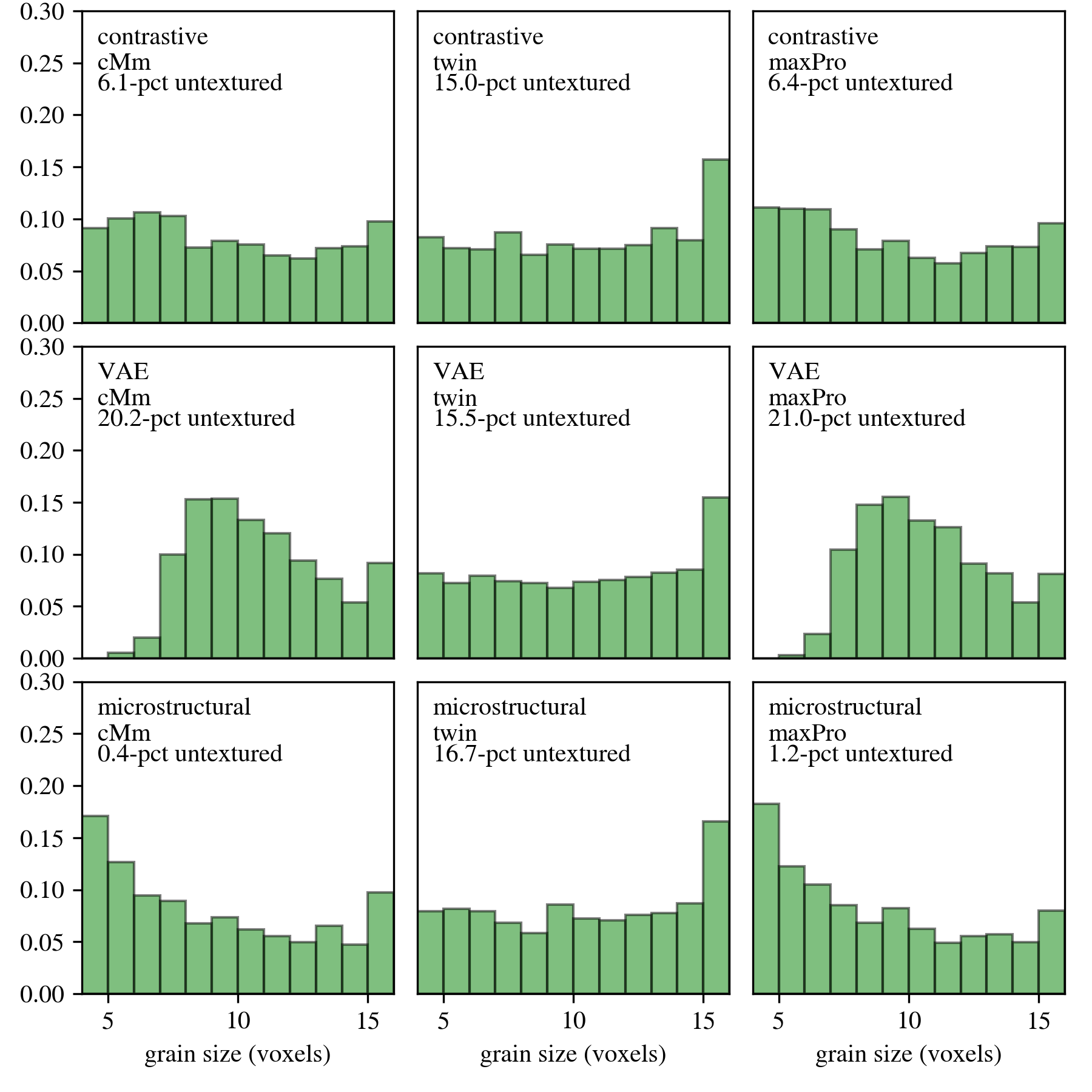}
  \caption{Distribution of grain sizes selected by each combination of design (cMm, maxPro, twin) and microstructural descriptors (contrastive, VAE, microstructural).}
  \label{fig:designDistros}
\end{center}
\end{figure}

\begin{figure}[]
\begin{center}
  \includegraphics[width=1.0\linewidth]{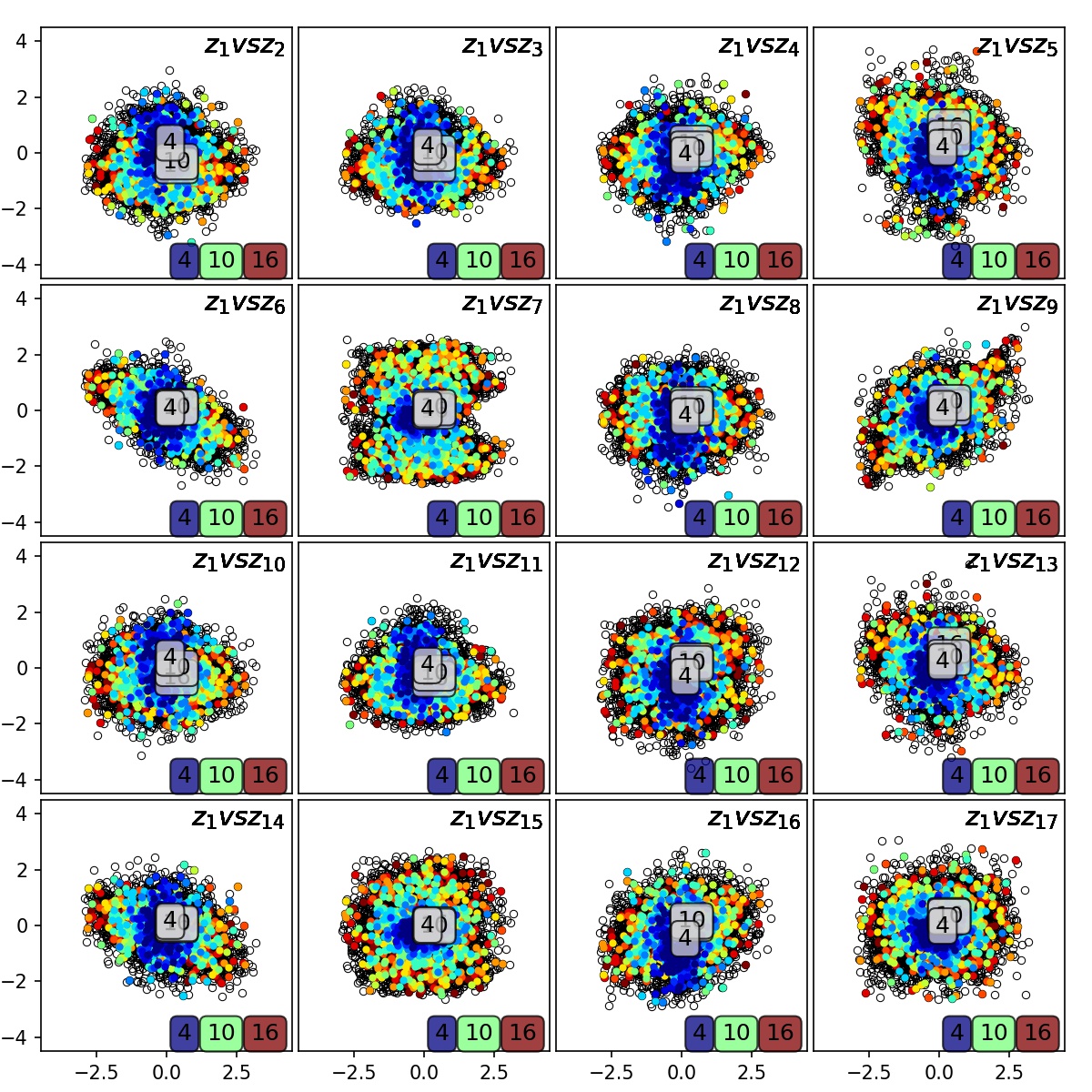}
  \caption{Latent space projections produced by the VAE. All data is shown but then data corresponding to uniformly textured material is filled with colors corresponding to grain size (centroid also includes labels 4, 10, 16).}
  \label{fig:latentSpaceVAE_0}
\end{center}
\end{figure}

In Fig. \ref{fig:designDistros} grain size histograms are shown for designs constructed using the three microstructure feature extraction methods (VAE, contrastive, microstructural) and three design criteria (cMm, maxPro, twin). Interestingly, across all designs contrastive features produce nearly uniform grain size distributions. cMm and maxPro designs with microstructural features seem to be prefer smaller grain sizes and neglect MVEs with preferred crystallographic texture. Conversley, cMm and maxPro designs on the VAE features are heavily weighted to neglect small grain size MVEs. We suspect that this disparity may be due to the dimensionality of the problem; contrastive features are 16-dimensional, microstructural 18-dimensional, and VAE features 512-dimensional. The latter was essential as the VAE is voxel-by-voxel trained to reconstruct an input MVE from the latent space. Good reconstruction performance, which needs to capture localized features throughout the volume, necessitated a high dimensional latent space. Hence, for designs such as cMm and maxPro, which seek to ``push'' points away from another, larger dimensionality representations will have a tendency to push points towards boundaries. In Fig. \ref{fig:latentSpaceVAE_0} several two-dimensional projections of the VAE latent space are shown. Markers are colored according to their grain size. Visually it appears that small grain MVEs are represented closer to the original in the latent space. This explains why both cMm and maxPro designs, which are designed to ``push'' points towards boundaries, would under-represent fine grain sizes. The twin design, however, balances both a space-filling objective and a distributional objective so that the design exhibits similar statistical properties as the full data set. Hence, the twin design criteria does not neglect fine grain sizes and proportionally represents untextured/textured examples.

\begin{figure}[]
\begin{center}
  \includegraphics[width=1.0\linewidth]{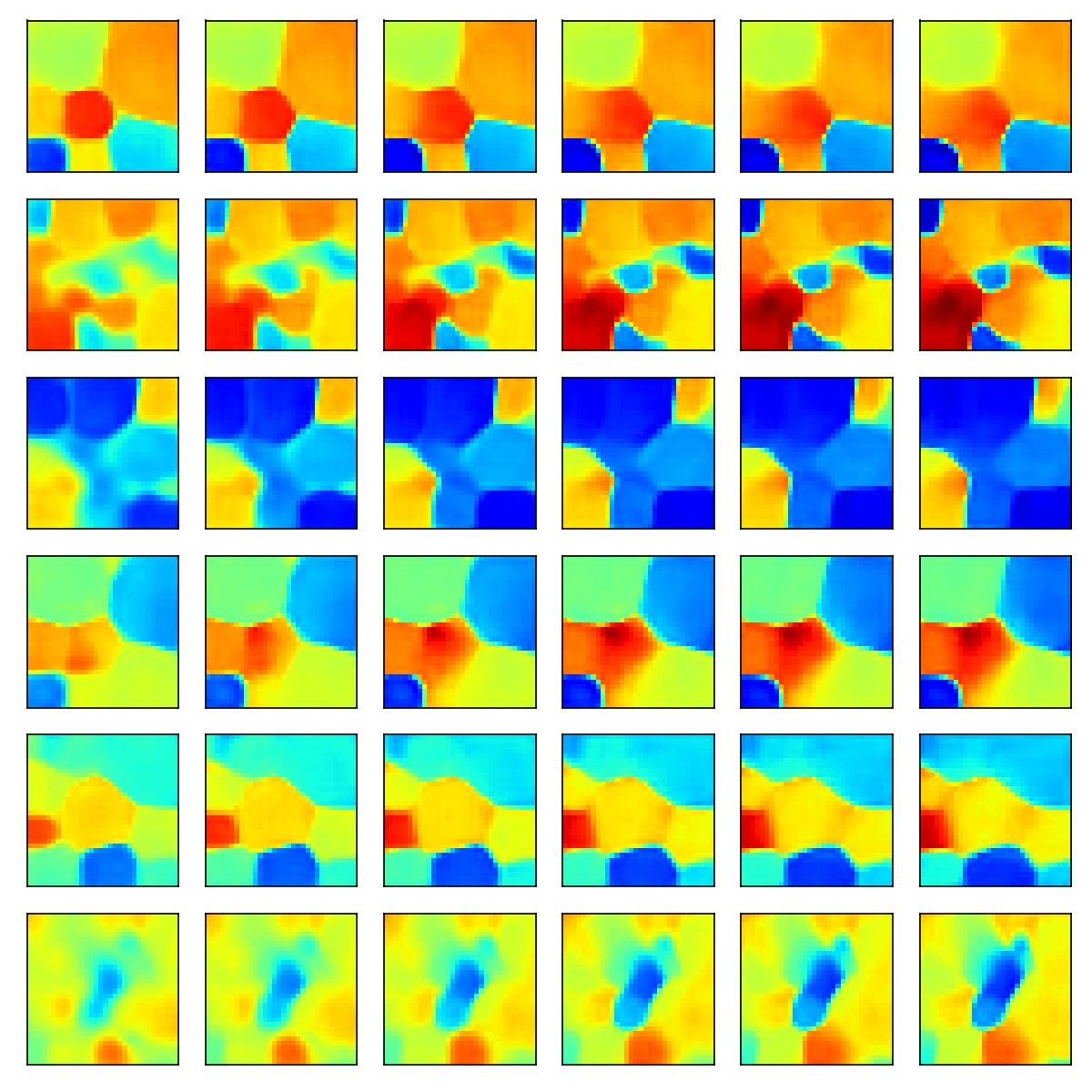}
  \caption{First column represents a two dimensional slice of a reconstructed $\hat{\bm{x}}$ from the validation data set. All other columns are interpolations from the latent space corresponding to $\alpha \cdot \bm{z}$ where $\alpha$ was varied from 1.2 to 2 linearly. These results indicate that the latent representation is indeed continuous and captures localized microstructural features.}
  \label{fig:latentSpacePerturbationPub_0}
\end{center}
\end{figure}

An important diagnostic to assess the VAE's behavior is to test the continuity of the latent space. Recall that in a VAE there is both a reconstruction loss and a KL term which encourages the latent space to be Gaussian and continuous. This is important since designs are being constructed using latent distances and, hence, obtaining a latent space where distances are meaningful is critical. Shown in Fig. \ref{fig:latentSpacePerturbationPub_0} are two dimensional reconstructions from the latent space. The first column corresponds to a reconstruction of examples from the validation data set. Each column after that is a reconstruction obtained from a linear mapping of the latent vector $\left\{1.2\cdot \bm{z},\ldots,2.0\cdot \bm{z}\right\}$. Each row is a different example. These figures confirm that indeed the VAE latent space is continuous with localized features being continuously manipulatable via perturbation of the latent representation.

\begin{figure}[]
\begin{center}
  \includegraphics[width=1.0\linewidth]{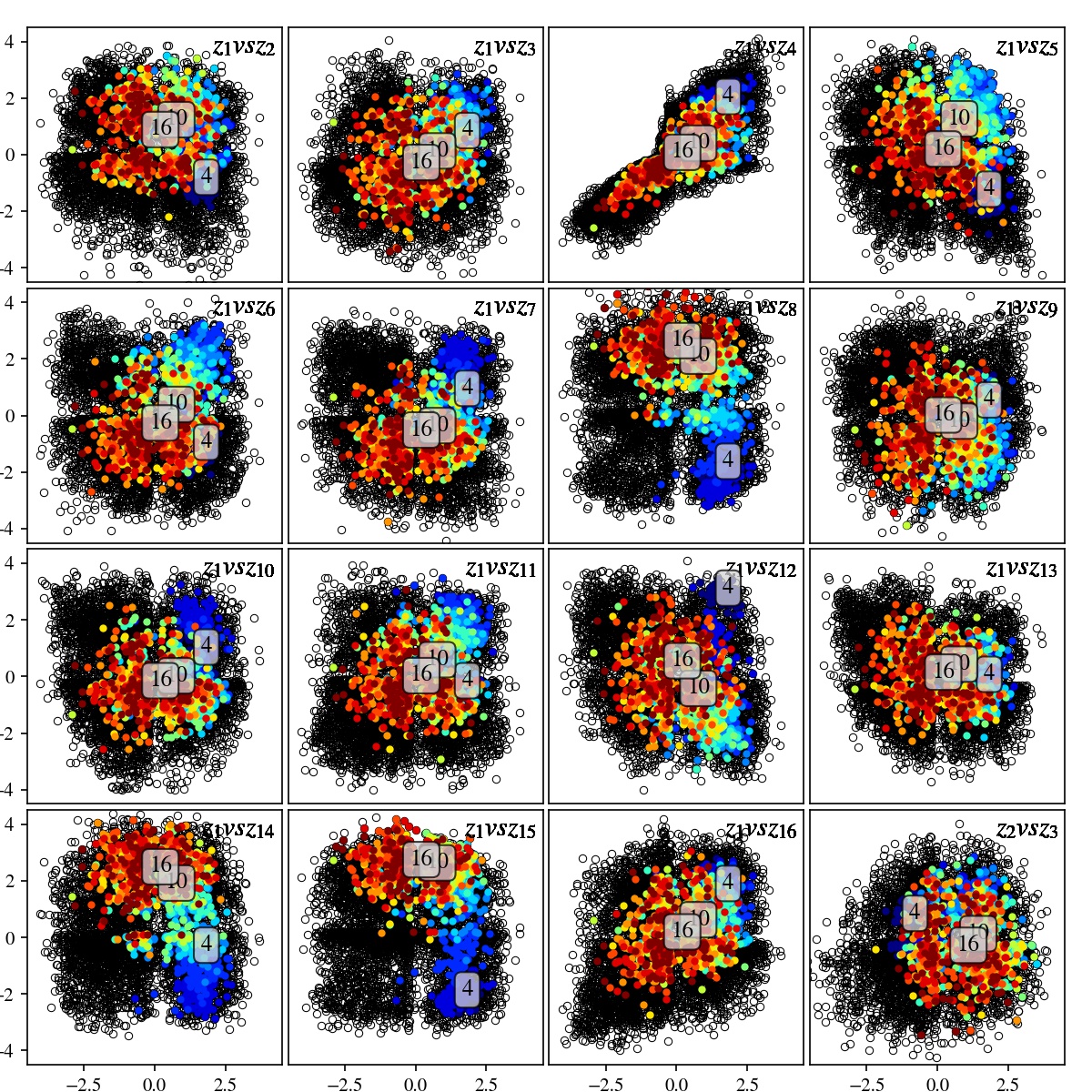}
  \caption{Latent space projections of the self-supervised microstructure statistics model. All data is shown but then data corresponding to uniformly textured material is filled with colors corresponding to grain size (centroid also includes labels 4, 10, 16). Clear structure is observed in the latent space.}
  \label{fig:latentSpace_0}
\end{center}
\end{figure}

\begin{figure}[]
\begin{center}
  \includegraphics[width=1.0\linewidth]{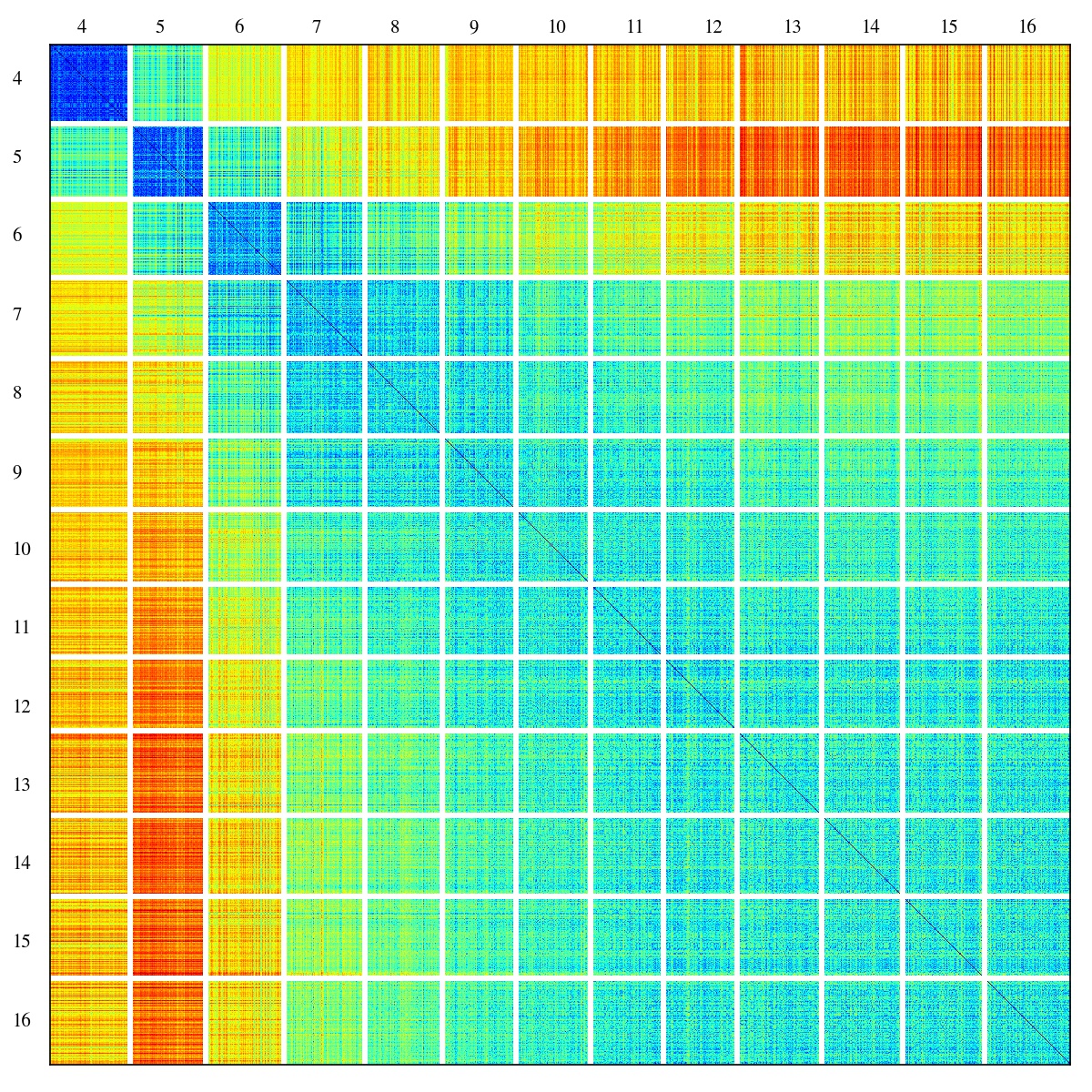}
  \caption{Distance matrix considering only uniformly random textured MVEs and comparing across grain sizes from 4 to 16 voxels.}
  \label{fig:distMatrixRandom_0}
\end{center}
\end{figure}

The contrastive latent space, shown in Fig. \ref{fig:latentSpace_0}, is by comparison drastically different. All data is shown but untextured examples are colored according to grain size. Visually it is clear that the latent space representation exhibits strong structural patterns with respect to both crystallographic texture and grain size; some regions are exclusively for untextured and there is a continuous gradation of grain size. The distance matrix for all untextured examples, shown in Fig. \ref{fig:distMatrixRandom_0}, further demonstrates the networks ability to discriminate across grain sizes. Note that for large grain sizes, about 12 voxels and above, the network cannot discriminate. We suspect that this is because the $16^3$ cropped example fed to the contrastive network is no longer representative for relatively large grain sizes. For small grain sizes, for instance the block corresponding to all 4-on-4 comparisons, distances are small indicating that the network clusters all examples together suggesting that it is indeed learning statistical descriptors.

\begin{figure}[]
\begin{center}
  \includegraphics[width=1.0\linewidth]{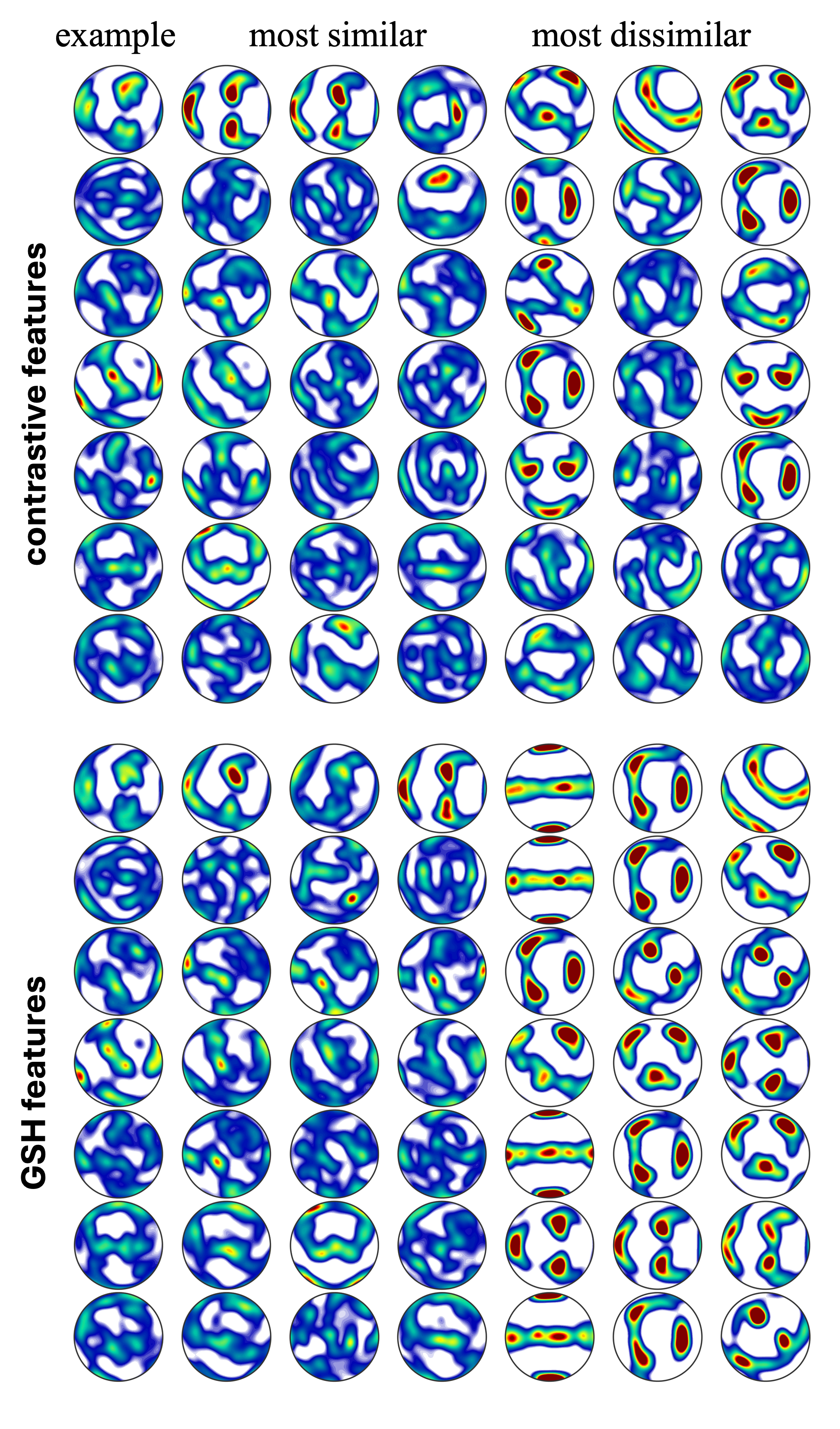}
  \caption{$(100)$ pole figures corresponding to: random example from the data set, the most similar instances, and most dissimilar instances. Similarity is measured using both the contrastive latent vector and GSH features. Identical color bar limits are used throughout.}
  \label{fig:poleFigures}
\end{center}
\end{figure}

The crystallographic texture sensitivity of the contrastive features is shown in Fig. \ref{fig:poleFigures}. Here random examples are sampled from the validation data set and then three of the closest and three of the most distant MVEs are identified using latent contrastive distances. Only data corresponding to examples with a grain size of 4 voxels was considered. The same exercise is performed using \textit{only} GSH coefficients for comparison. Results indicate that the contrastive features do indeed capture texture similarity accurately. In a few instances there are visual anomalies but it is important to consider that the GSH features only capture texture whereas the contrastive features capture both texture and structural information. Nonetheless, these results indicate that the established self-supervised feature extraction network does capture both spatial and crystallographic features automatically.

\subsection{Surrogate training results}
\label{section:surrogateResults}

\begin{figure}[]
\begin{center}
  \includegraphics[width=1.0\linewidth]{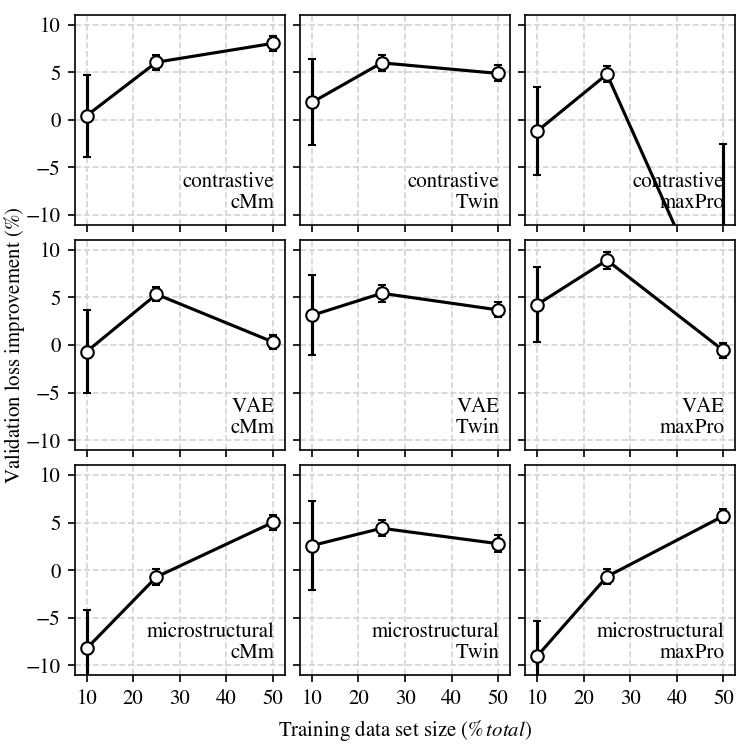}
  \caption{Parametric results for all considered microstructural features and design criteria. Validation loss improvement is compared against models trained from randomly selected training designs. For each feature/design combination 10 designs were used to train 10 models. Error bars correspond to one standard deviation computed via bootstrapping.}
  \label{fig:summaryResults}
\end{center}
\end{figure}

Summary results for the MVE-feature/design parametric study are shown in Fig. \ref{fig:summaryResults}. Across all possible combinations the VAE-maxPro combination scored the highest improvement at 8.8\% for the case where only 25\% of the total data set size was used. cMm with contrastive features appears to be systematically most robust. Interestingly the performance of the VAE-maxPro combination deteriorates completely when 50\% of the total data set size was used. Similarly, the contrastive/maxPro combination also suffers a significant deterioration at 50\% data set use, however, microstructural features still enjoy a moderate boost. We suspect that this may be due to the difficulty in optimizing designs. Recall that the design criteria is represented by a scalar valued objective function; constructing the design requires solving a high dimensional optimization problem. The maxPro objective function, Eqn. \ref{eqn:maxPro}, is rather challenging to optimize as it contains $n^2-n$ terms with each term a product of $p$ terms. In the VAE case $p=512$ and for 50\% ($n=3413$) there are over 11-million terms. Note that some of this complexity is reduced by using a greedy optimization strategy but, nonetheless, the optimization problem remains challenging. However, the contrastive ($p=16$) and microstructural ($p=18$) roughly are of the same dimension and yet only the contrastive features deteriorate at 50\% data set size using the maxPro criteria. It may be possible that this is because the contrastive structural-orientation features are coupled whereas for the constructed microstructural feature vector they are not. The maxPro criteria considers distances in projected spaces which will make optimization more challenging if features are entangled. The decoupled (grain size and crystallographic texture) microstructural features are less susceptible to this effect.

Somewhat remarkably the data twinning design performed nearly identically across all considered MVE features with a nearly constant 5\% boost. This may be because the design criteria has a balanced objective function which also considers the underlying data distribution. If the problem dimensionality is large, and the desired design size small, then cMm and maxPro can have a tendency to push points towards extreme ``corners'' producing designs which perform poorer than random designs. It is suspected that the data twinning designs mitigate against this by penalizing designs which do not emulate the original data distribution. This also may  be why a slight decline in performance is observed at 50\%; eventually random sampling will begin to represent the underlying data distribution and, hence, the twinning design's benefit will decay.

Finally, consider that for all designs-feature combinations, with the exception of maxPro/cMm and microstructural features, there is no statistically significant boost at 10\% of the total data set size. However, for cMm/maxPro designs and microstructural features there is a statistically significant \textit{decline} in performance relative to random designs. Previously it was hypothesized that for large data set sizes the entangling of features may make optimization of the design criteria more difficult. In the case of small data set sizes we argue that the converse may hold true; entanglement aids in avoiding ``corner biased'' designs. When features are independent the pace-filling criteria may bias the data set towards certain grain size and texture corners of the 18-dimensional microstructural feature space. Increases in the data set size remedy this behavior and result in monotonically increasing performance.

\begin{figure}[]
\begin{center}
  \includegraphics[width=1.0\linewidth]{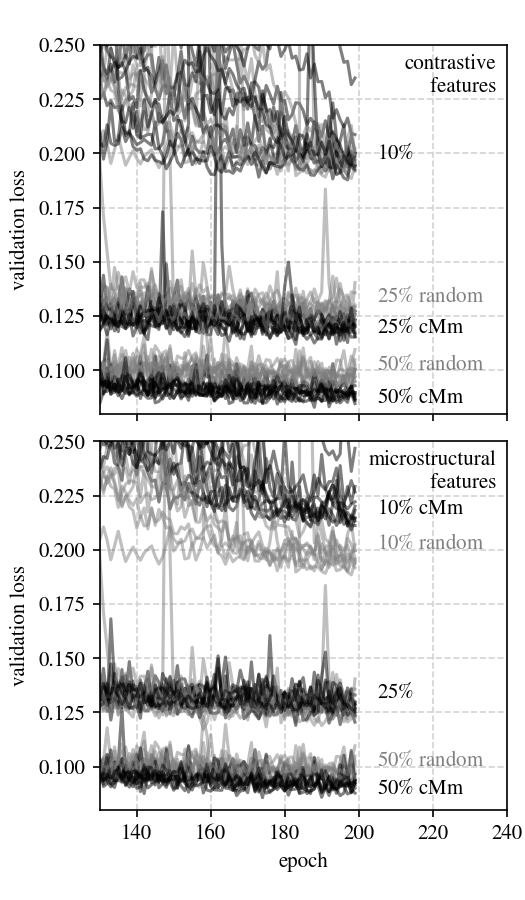}
  \caption{Validation loss curves for a few selected designs and features. All ten curves shown to demonstrate repeatability.}
  \label{fig:someLosses}
\end{center}
\end{figure}

Validation loss curves for a few select feature and design combinations are shown in Fig. \ref{fig:someLosses}. Note that the loss here corresponds to the mean of centered and normalized components of the stress tensor e.g. the value is a non-dimensional quantity. Visually it is clear that indeed at certain data set sizes there appears a significant improvement in surrogate model performance as summarized in Fig. \ref{fig:summaryResults}. The loss curves at 10\% data set size reveal that the loss curves appear to be somewhat unstable exhibiting a great deal of variance. This reveals that for the specific architecture used the 10\% (about 600 simulations) size may be at the very limit of the models training stability which may explain some of the previously discussed anomalous results (worse than random design results at 10\%).

\section{Discussion}
\label{section:discussion}

\begin{figure}[]
\begin{center}
  \includegraphics[width=1.0\linewidth]{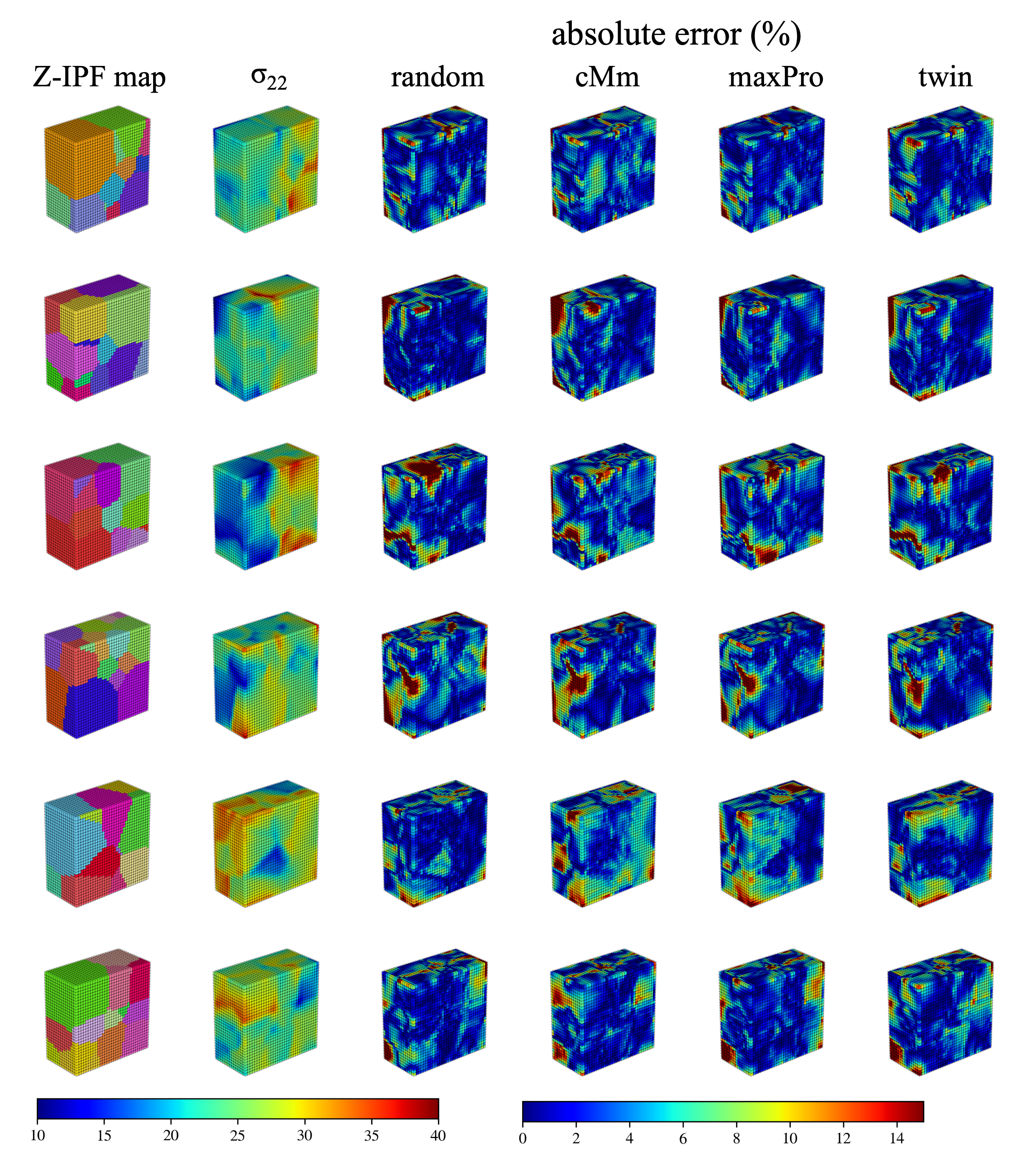}
  \caption{Random examples from validation data: IPF map, $\sigma_{33}$ of the stress tensor, and absolute error maps for four designs (using contrastive features and 25\% of the dataset). Examples are selected from one of the ten surrogate realizations with validation loss closest to the mean validation loss.}
  \label{fig:absError}
\end{center}
\end{figure}

\begin{figure}[]
\begin{center}
  \includegraphics[width=1.0\linewidth]{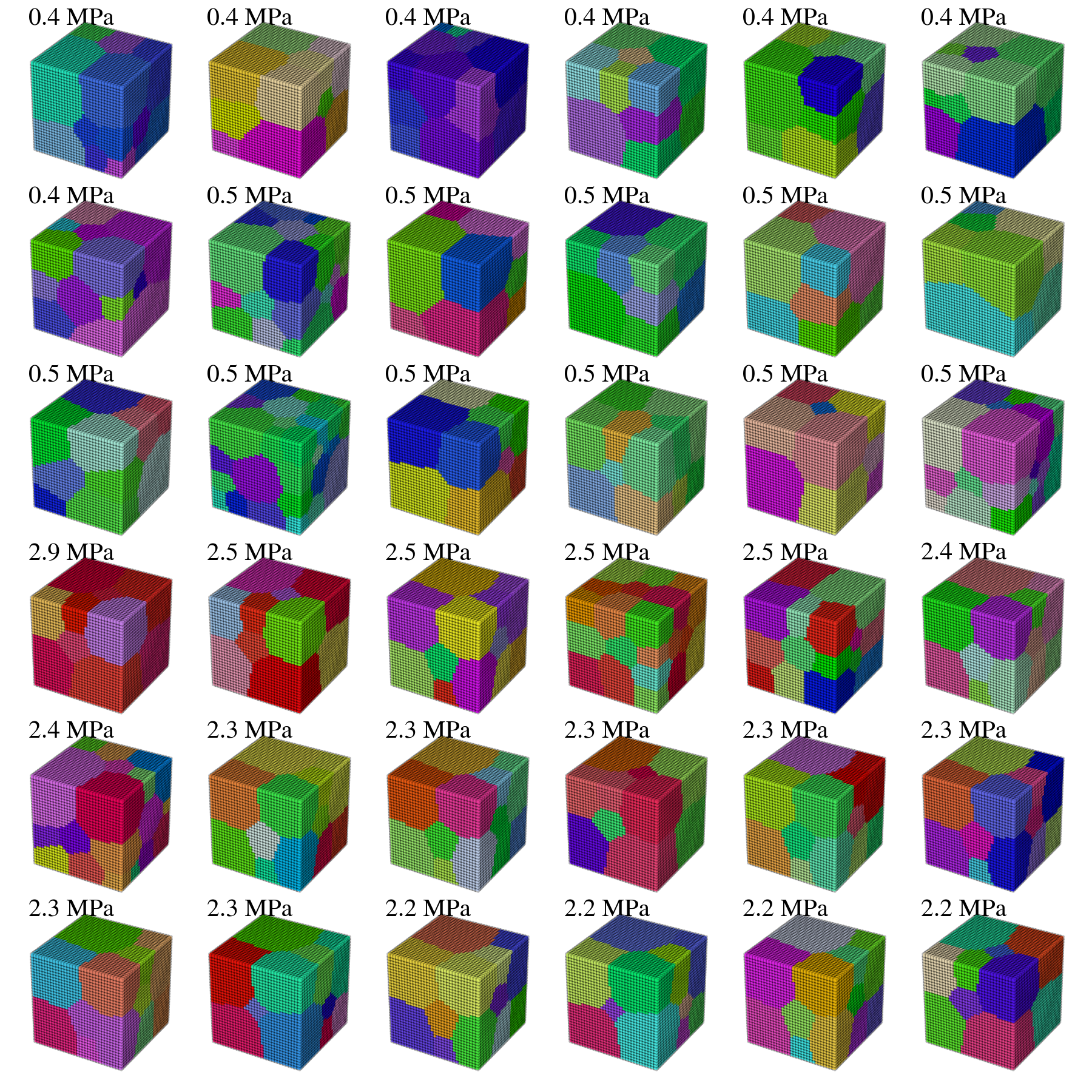}
  \caption{Example microstructures with best (top) and worst (bottom) volume averaged error. Provided number is the volume averaged $\sigma_{33}$ standard error.}
  \label{fig:goodBadStructures}
\end{center}
\end{figure}

A number of validation example MVEs, FE results, and surrogate results are shown in Fig. \ref{fig:absError}. For visual comparison surrogate fields are shown as absolute percent error relative to the FE results. While there are subtle differences in responses across the various designs these results do not immediately shed any light on the impact of features and the design of surrogate performance. To further explore any potential insights the best 18 and worst 18 performance MVEs from a contrastive cMm with 25\% data are shown in Fig. \ref{fig:goodBadStructures}. Remarkably, the best and worst performing structures all visually appear to be extremely similar.

\begin{figure}[]
\begin{center}
  \includegraphics[width=1.0\linewidth]{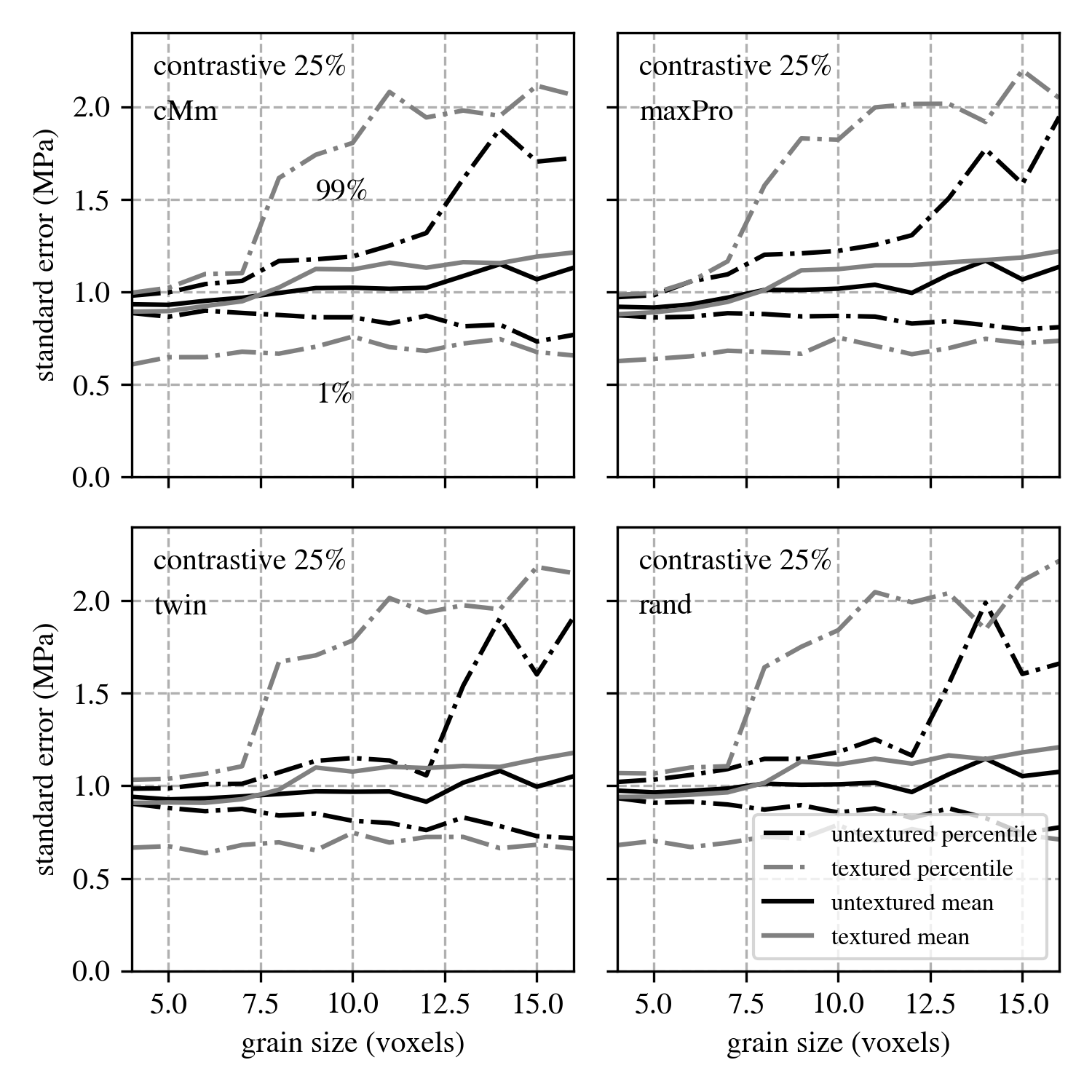}
  \caption{1 and 99 percentiles and mean volume-averaged standard error for various designs using contrastive features and 25\% of the data set. Data is shown as a function of grain size and texture/untextured. Results indicate that while mean behavior is similar across all microstructures large grain textured examples tend to exhibit a fatter tail of predictive capacity.}
  \label{fig:errorStats}
\end{center}
\end{figure}

Surrogate model performance trends as a function of microstructural features are shown in Fig. \ref{fig:errorStats}. The mean standard error 1 and 99-percentile bounds are shown for all grain sizes and textured/untextured structures. It is clear that for all design criteria there are far more extreme value prediction MVEs exhibiting both very large and small errors at large grain sizes. This indicates that the ``tails'' of the surrogates' performance are much wider for large grain sizes. It also appears that this effect is even more pronounced for textured materials. A possible explanation is that for small grain size the MVEs are behaving as representative volume elements (RVEs) which capture aggregate material behavior well. Furthermore, RVEs may intrinsically contain more information; a single volume contains hundreds of pairs of small grain neighbor combinations. In large grain size MVEs each example will contain far fewer of these neighbor pairs. This is important because the ML surrogate must be exposed to a diversity of structures to effectively learn key physical relationships. Texture adds another degree of complexity that results in more extreme value performance responses. Not only do spatial neighborhood relationships need to be learned but also the response as a function of specific aggregate crystallographic textures.

The presented results demonstrate that micromechanical surrogate models trained using well designed MVEs can enjoy a moderate boost in overall performance when compared against randomly selected training MVEs. However, the parametric study over several MVE feature descriptors and design criteria revealed that net performance is dependent on many factors and these are difficult to assess a-priori. For the elastic constitutive model considered here the ``cost'' of this uncertainty is rather low because evaluation of the FE model is computationally rather inexpensive. However, for more advanced inelastic constitutive models significant computational cost is incurred when evaluating the model and, hence, it is imperative to identify designs that will not result in worse than random designs.

A key consideration when constructing designs is the optimization complexity for a specific data set size and problem dimensionality. Results indicated that while maxPro performed well for 25\% data set sizes for both VAE and contrastive features, performance deteriorated completely at 50\% data set size resulting in worse or no performance over random designs. Our hypothesis is that this is because construction of a design requires solving an optimization problem and, hence, the bottleneck in the process is the design optimization step. In the VAE case both the relatively small data set size and the dimensionality of the problem ($p=512$) cause issues. For problems with much larger data set sizes some of these issues may be alleviated.

While the cMm design is inferior to maxPro with respect to projection quality the cMm objective is much more easy to optimize. Furthermore, optimization is even more feasible for small MVE latent state representations. This is likely why the cMm-contrastive approach yielded consistently good results (monotonically increasing performance) with increasing data set size. While the microstructural-cMm combination also yielded monotonically increasing performance in the case where surrogates were trained used 10\% of the data set size these designs yielded worse than random performance. It is suspected that this may be because of the decoupled nature of the microstructural features (grain size and volume averaged texture). Contrastive features are capable of capturing richer coupled richer features which may offer opportunity for ensuring diversity. Finally, the data twinning approach, while robust against poorer than random performance, only provides moderate boosts for moderate data set sizes. The robustness is believed to be due to the tendency to construct designs to emulate the original data set's distribution and, hence, mitigate against designs dominated by anomalies e.g. clustering in high dimensional corners. However, this comes at the cost of locating points close to one another which minimizes overall diversity. 

Based on these results, for the specific surrogate architecture and micromechanical model considered, the most promising feature/design combination is the contrastive conditional maximin design. While a rather moderate 8\% boost in validation performance was achieved this may be even better for other domain problems and larger data set sizes. Consider a case where instead of 6,825 MVE candidate structures are available there are instead 68,250 and a total budget of 20,000 simulations to be performed. The larger candidate set size would provide more opportunity to diversify the design and a more expansive data set would allow for more efficient packing of the feature space. Furthermore, based on the results from Fig. \ref{fig:errorStats} it is clear that more examples are needed for the less representative large grain textured MVEs. Finally, it should be noted much of these results

\section{Conclusions}
\label{section:conclusions}

The proposed feature-extraction and experimental design strategy for establishing micromechanics surrogates has been demonstrated to be effective. The strategy consists of two key components: (1) a microstructural feature for computing a distance or similarity metric and (2) a design strategy for distributing points in the microstructural feature space. A parametric study was performed over three design strategies and three microstructural features. Results show that for the considered problem a reduction up to 8\% of the validation loss is achieved when compared to random selection of training examples. Trends indicate that for bigger data sets the benefits may be even larger. This is rationalized by the notion that high-dimensional spaces are difficult to fill, and hence, larger designs will facilitate more efficient sampling of the microstructural feature space. This in turn ensures a more diverse training data set which greatly improves model performance. In addition to demonstrating this result this work also establishes a novel self-supervised contrastive feature extraction methodology for computing microstructural summary statistics.

\section{Data availability}
\label{section:data}

Data will be made available from the authors upon reasonable request.

\section*{Acknowledgements}

Research was sponsored by the US Department of Energy, Office of Energy Efficiency and Renewable Energy (EERE), Advanced Manufacturing Office, and Advanced Materials and Manufacturing Technologies Office (AMMTO) under contract DE-AC05-00OR22725 with UT-Battelle LLC and performed in partiality at the Oak Ridge National Laboratory’s Manufacturing Demonstration Facility, an Office of Energy Efficiency and Renewable Energy user facility. All the authors would like to acknowledge the support of the HPC4Mtls program. 
 
\FloatBarrier
\label{references}
\bibliographystyle{unsrtnat}
\bibliography{refs}

\begin{thebibliography}{53}
\providecommand{\natexlab}[1]{#1}
\providecommand{\url}[1]{\texttt{#1}}
\expandafter\ifx\csname urlstyle\endcsname\relax
  \providecommand{\doi}[1]{doi: #1}\else
  \providecommand{\doi}{doi: \begingroup \urlstyle{rm}\Url}\fi

\bibitem[De~Pablo et~al.(2014)De~Pablo, Jones, Kovacs, Ozolins, and Ramirez]{de2014materials}
Juan~J De~Pablo, Barbara Jones, Cora~Lind Kovacs, Vidvuds Ozolins, and Arthur~P Ramirez.
\newblock The materials genome initiative, the interplay of experiment, theory and computation.
\newblock \emph{Current Opinion in Solid State and Materials Science}, 18\penalty0 (2):\penalty0 99--117, 2014.

\bibitem[Torquato and Haslach~Jr(2002)]{torquato2002random}
Salvatore Torquato and Henry~W Haslach~Jr.
\newblock Random heterogeneous materials: microstructure and macroscopic properties.
\newblock \emph{Appl. Mech. Rev.}, 55\penalty0 (4):\penalty0 B62--B63, 2002.

\bibitem[Fernandez-Zelaia et~al.(2024)Fernandez-Zelaia, Cheng, Mayeur, Ziabari, and Kirka]{fernandez2024digital}
Patxi Fernandez-Zelaia, Jiahao Cheng, Jason Mayeur, Amir~Koushyar Ziabari, and Michael~M Kirka.
\newblock Digital polycrystalline microstructure generation using diffusion probabilistic models.
\newblock \emph{Materialia}, 33:\penalty0 101976, 2024.

\bibitem[Buzzy et~al.(2024)Buzzy, Robertson, and Kalidindi]{buzzy2024statistically}
Michael~O Buzzy, Andreas~E Robertson, and Surya~R Kalidindi.
\newblock Statistically conditioned polycrystal generation using denoising diffusion models.
\newblock \emph{Acta Materialia}, page 119746, 2024.

\bibitem[Robertson et~al.(2023)Robertson, Kelly, Buzzy, and Kalidindi]{robertson2023local}
Andreas~E Robertson, Conlain Kelly, Michael Buzzy, and Surya~R Kalidindi.
\newblock Local--global decompositions for conditional microstructure generation.
\newblock \emph{Acta Materialia}, 253:\penalty0 118966, 2023.

\bibitem[Lee and Yun(2023)]{lee2023microstructure}
Kang-Hyun Lee and Gun~Jin Yun.
\newblock Microstructure reconstruction using diffusion-based generative models.
\newblock \emph{Mechanics of Advanced Materials and Structures}, pages 1--19, 2023.

\bibitem[Zeni et~al.(2023)Zeni, Pinsler, Z{\"u}gner, Fowler, Horton, Fu, Shysheya, Crabb{\'e}, Sun, Smith, et~al.]{zeni2023mattergen}
Claudio Zeni, Robert Pinsler, Daniel Z{\"u}gner, Andrew Fowler, Matthew Horton, Xiang Fu, Sasha Shysheya, Jonathan Crabb{\'e}, Lixin Sun, Jake Smith, et~al.
\newblock Mattergen: a generative model for inorganic materials design.
\newblock \emph{arXiv preprint arXiv:2312.03687}, 2023.

\bibitem[Yang et~al.(2023)Yang, Cho, Merchant, Abbeel, Schuurmans, Mordatch, and Cubuk]{yang2023scalable}
Mengjiao Yang, KwangHwan Cho, Amil Merchant, Pieter Abbeel, Dale Schuurmans, Igor Mordatch, and Ekin~Dogus Cubuk.
\newblock Scalable diffusion for materials generation.
\newblock \emph{arXiv preprint arXiv:2311.09235}, 2023.

\bibitem[Sacks et~al.(1989)Sacks, Schiller, and Welch]{sacks1989designs}
Jerome Sacks, Susannah~B Schiller, and William~J Welch.
\newblock Designs for computer experiments.
\newblock \emph{Technometrics}, 31\penalty0 (1):\penalty0 41--47, 1989.

\bibitem[Kennedy and O'Hagan(2001)]{kennedy2001bayesian}
Marc~C Kennedy and Anthony O'Hagan.
\newblock Bayesian calibration of computer models.
\newblock \emph{Journal of the Royal Statistical Society: Series B (Statistical Methodology)}, 63\penalty0 (3):\penalty0 425--464, 2001.

\bibitem[Gramacy(2020)]{gramacy2020surrogates}
Robert~B Gramacy.
\newblock \emph{Surrogates: Gaussian process modeling, design, and optimization for the applied sciences}.
\newblock Chapman and Hall/CRC, 2020.

\bibitem[Mak et~al.(2018)Mak, Sung, Wang, Yeh, Chang, Joseph, Yang, and Wu]{mak2018efficient}
Simon Mak, Chih-Li Sung, Xingjian Wang, Shiang-Ting Yeh, Yu-Hung Chang, V~Roshan Joseph, Vigor Yang, and CF~Jeff Wu.
\newblock An efficient surrogate model for emulation and physics extraction of large eddy simulations.
\newblock \emph{Journal of the American Statistical Association}, 113\penalty0 (524):\penalty0 1443--1456, 2018.

\bibitem[Chen et~al.(2021)Chen, Mak, Joseph, and Zhang]{chen2021function}
Jialei Chen, Simon Mak, V~Roshan Joseph, and Chuck Zhang.
\newblock Function-on-function kriging, with applications to three-dimensional printing of aortic tissues.
\newblock \emph{Technometrics}, 63\penalty0 (3):\penalty0 384--395, 2021.

\bibitem[Damianou and Lawrence(2013)]{damianou2013deep}
Andreas Damianou and Neil~D Lawrence.
\newblock Deep gaussian processes.
\newblock In \emph{Artificial intelligence and statistics}, pages 207--215. PMLR, 2013.

\bibitem[Dai et~al.(2015)Dai, Damianou, Gonz{\'a}lez, and Lawrence]{dai2015variational}
Zhenwen Dai, Andreas Damianou, Javier Gonz{\'a}lez, and Neil Lawrence.
\newblock Variational auto-encoded deep gaussian processes.
\newblock \emph{arXiv preprint arXiv:1511.06455}, 2015.

\bibitem[Tran and Wildey(2021)]{tran2021solving}
Anh Tran and Tim Wildey.
\newblock Solving stochastic inverse problems for property--structure linkages using data-consistent inversion and machine learning.
\newblock \emph{JOM}, 73\penalty0 (1):\penalty0 72--89, 2021.

\bibitem[Liang et~al.(2018)Liang, Liu, Martin, and Sun]{liang2018deep}
Liang Liang, Minliang Liu, Caitlin Martin, and Wei Sun.
\newblock A deep learning approach to estimate stress distribution: a fast and accurate surrogate of finite-element analysis.
\newblock \emph{Journal of The Royal Society Interface}, 15\penalty0 (138):\penalty0 20170844, 2018.

\bibitem[He et~al.(2023)He, Koric, Kushwaha, Park, Abueidda, and Jasiuk]{he2023novel}
Junyan He, Seid Koric, Shashank Kushwaha, Jaewan Park, Diab Abueidda, and Iwona Jasiuk.
\newblock Novel deeponet architecture to predict stresses in elastoplastic structures with variable complex geometries and loads.
\newblock \emph{Computer Methods in Applied Mechanics and Engineering}, 415:\penalty0 116277, 2023.

\bibitem[Montes~de Oca~Zapiain et~al.(2022{\natexlab{a}})Montes~de Oca~Zapiain, Shanker, and Kalidindi]{montes2022convolutional}
David Montes~de Oca~Zapiain, Apaar Shanker, and Surya~R Kalidindi.
\newblock Convolutional neural networks for the localization of plastic velocity gradient tensor in polycrystalline microstructures.
\newblock \emph{Journal of Engineering Materials and Technology}, 144\penalty0 (1):\penalty0 011004, 2022{\natexlab{a}}.

\bibitem[Yabansu et~al.(2014)Yabansu, Patel, and Kalidindi]{yabansu2014calibrated}
Yuksel~C Yabansu, Dipen~K Patel, and Surya~R Kalidindi.
\newblock Calibrated localization relationships for elastic response of polycrystalline aggregates.
\newblock \emph{Acta Materialia}, 81:\penalty0 151--160, 2014.

\bibitem[Khorrami et~al.(2023)Khorrami, Mianroodi, Siboni, Goyal, Svendsen, Benner, and Raabe]{khorrami2023artificial}
Mohammad~S Khorrami, Jaber~R Mianroodi, Nima~H Siboni, Pawan Goyal, Bob Svendsen, Peter Benner, and Dierk Raabe.
\newblock An artificial neural network for surrogate modeling of stress fields in viscoplastic polycrystalline materials.
\newblock \emph{npj Computational Materials}, 9\penalty0 (1):\penalty0 37, 2023.

\bibitem[Wang et~al.(2021)Wang, Oyen, Guo, Mehta, Scott, Panda, Fern{\'a}ndez-Godino, Srinivasan, and Yue]{wang2021stressnet}
Yinan Wang, Diane Oyen, Weihong Guo, Anishi Mehta, Cory~Braker Scott, Nishant Panda, M~Giselle Fern{\'a}ndez-Godino, Gowri Srinivasan, and Xiaowei Yue.
\newblock Stressnet-deep learning to predict stress with fracture propagation in brittle materials.
\newblock \emph{Npj Materials Degradation}, 5\penalty0 (1):\penalty0 6, 2021.

\bibitem[Pokharel et~al.(2021)Pokharel, Pandey, and Scheinker]{pokharel2021physics}
Reeju Pokharel, Anup Pandey, and Alexander Scheinker.
\newblock Physics-informed data-driven surrogate modeling for full-field 3d microstructure and micromechanical field evolution of polycrystalline materials.
\newblock \emph{JOM}, 73\penalty0 (11):\penalty0 3371--3382, 2021.

\bibitem[Pandey and Pokharel(2021)]{pandey2021machine}
Anup Pandey and Reeju Pokharel.
\newblock Machine learning based surrogate modeling approach for mapping crystal deformation in three dimensions.
\newblock \emph{Scripta Materialia}, 193:\penalty0 1--5, 2021.

\bibitem[Wu and Hamada(2011)]{wu2011experiments}
CF~Jeff Wu and Michael~S Hamada.
\newblock \emph{Experiments: planning, analysis, and optimization}.
\newblock John Wiley \& Sons, 2011.

\bibitem[Joseph(2016)]{joseph2016space}
V~Roshan Joseph.
\newblock Space-filling designs for computer experiments: A review.
\newblock \emph{Quality Engineering}, 28\penalty0 (1):\penalty0 28--35, 2016.

\bibitem[Bessa et~al.(2017)Bessa, Bostanabad, Liu, Hu, Apley, Brinson, Chen, and Liu]{bessa2017framework}
Miguel~A Bessa, Ramin Bostanabad, Zeliang Liu, Anqi Hu, Daniel~W Apley, Catherine Brinson, Wei Chen, and Wing~Kam Liu.
\newblock A framework for data-driven analysis of materials under uncertainty: Countering the curse of dimensionality.
\newblock \emph{Computer Methods in Applied Mechanics and Engineering}, 320:\penalty0 633--667, 2017.

\bibitem[Paulson et~al.(2017)Paulson, Priddy, McDowell, and Kalidindi]{paulson2017reduced}
Noah~H Paulson, Matthew~W Priddy, David~L McDowell, and Surya~R Kalidindi.
\newblock Reduced-order structure-property linkages for polycrystalline microstructures based on 2-point statistics.
\newblock \emph{Acta Materialia}, 129:\penalty0 428--438, 2017.

\bibitem[He et~al.(2024)He, Pal, Najafi, Abueidda, Koric, and Jasiuk]{he2024material}
Junyan He, Deepankar Pal, Ali Najafi, Diab Abueidda, Seid Koric, and Iwona Jasiuk.
\newblock Material-response-informed deeponet and its application to polycrystal stress-strain prediction in crystal plasticity.
\newblock \emph{arXiv preprint arXiv:2401.09977}, 2024.

\bibitem[Montes~de Oca~Zapiain et~al.(2022{\natexlab{b}})Montes~de Oca~Zapiain, Wood, Lubbers, Pereyra, Thompson, and Perez]{montes2022training}
David Montes~de Oca~Zapiain, Mitchell~A Wood, Nicholas Lubbers, Carlos~Z Pereyra, Aidan~P Thompson, and Danny Perez.
\newblock Training data selection for accuracy and transferability of interatomic potentials.
\newblock \emph{npj Computational Materials}, 8\penalty0 (1):\penalty0 189, 2022{\natexlab{b}}.

\bibitem[Barneschi et~al.(2024)Barneschi, Rotondi, and Padula]{barneschi2024molecular}
Leonardo Barneschi, Leonardo Rotondi, and Daniele Padula.
\newblock Molecular geometry impact on deep learning predictions of inverted singlet--triplet gaps.
\newblock \emph{The Journal of Physical Chemistry A}, 2024.

\bibitem[Sener and Savarese(2017)]{sener2017active}
Ozan Sener and Silvio Savarese.
\newblock Active learning for convolutional neural networks: A core-set approach.
\newblock \emph{arXiv preprint arXiv:1708.00489}, 2017.

\bibitem[Joseph et~al.(2015)Joseph, Gul, and Ba]{joseph2015maximum}
V~Roshan Joseph, Evren Gul, and Shan Ba.
\newblock Maximum projection designs for computer experiments.
\newblock \emph{Biometrika}, 102\penalty0 (2):\penalty0 371--380, 2015.

\bibitem[Vakayil and Joseph(2022)]{vakayil2022data}
Akhil Vakayil and V~Roshan Joseph.
\newblock Data twinning.
\newblock \emph{Statistical Analysis and Data Mining: The ASA Data Science Journal}, 15\penalty0 (5):\penalty0 598--610, 2022.

\bibitem[Yaghoobi et~al.(2019)Yaghoobi, Ganesan, Sundar, Lakshmanan, Rudraraju, Allison, and Sundararaghavan]{yaghoobi2019prisms}
Mohammadreza Yaghoobi, Sriram Ganesan, Srihari Sundar, Aaditya Lakshmanan, Shiva Rudraraju, John~E Allison, and Veera Sundararaghavan.
\newblock Prisms-plasticity: An open-source crystal plasticity finite element software.
\newblock \emph{Computational Materials Science}, 169:\penalty0 109078, 2019.

\bibitem[Fernandez-Zelaia et~al.(2022)Fernandez-Zelaia, Lee, Dryepondt, and Kirka]{fernandez2022creep}
Patxi Fernandez-Zelaia, Yousub Lee, Sebastien Dryepondt, and Michael~M Kirka.
\newblock Creep anisotropy modeling and uncertainty quantification of an additively manufactured ni-based superalloy.
\newblock \emph{International Journal of Plasticity}, 151:\penalty0 103177, 2022.

\bibitem[Quey and Kasemer(2022)]{quey2022neper}
Romain Quey and Matthew Kasemer.
\newblock The neper/fepx project: free/open-source polycrystal generation, deformation simulation, and post-processing.
\newblock In \emph{IOP Conference Series: Materials Science and Engineering}, volume 1249, page 012021. IOP Publishing, 2022.

\bibitem[Bunge(2013)]{bunge2013texture}
H-J Bunge.
\newblock \emph{Texture analysis in materials science: mathematical methods}.
\newblock Elsevier, 2013.

\bibitem[Goodfellow et~al.(2016)Goodfellow, Bengio, and Courville]{goodfellow2016deep}
Ian Goodfellow, Yoshua Bengio, and Aaron Courville.
\newblock \emph{Deep learning}.
\newblock MIT press, 2016.

\bibitem[Higgins et~al.(2017)Higgins, Matthey, Pal, Burgess, Glorot, Botvinick, Mohamed, and Lerchner]{higgins2017beta}
Irina Higgins, Loic Matthey, Arka Pal, Christopher~P Burgess, Xavier Glorot, Matthew~M Botvinick, Shakir Mohamed, and Alexander Lerchner.
\newblock beta-vae: Learning basic visual concepts with a constrained variational framework.
\newblock \emph{ICLR (Poster)}, 3, 2017.

\bibitem[Hashemi et~al.(2024)Hashemi, Guerzhoy, and Paulson]{hashemi2024toward}
Sayed~Sajad Hashemi, Michael Guerzhoy, and Noah~H Paulson.
\newblock Toward learning latent-variable representations of microstructures by optimizing in spatial statistics space.
\newblock \emph{arXiv preprint arXiv:2402.11103}, 2024.

\bibitem[Larroza et~al.(2016)Larroza, Bod{\'\i}, Moratal, et~al.]{larroza2016texture}
Andr{\'e}s Larroza, Vicente Bod{\'\i}, David Moratal, et~al.
\newblock Texture analysis in magnetic resonance imaging: review and considerations for future applications.
\newblock \emph{Assessment of cellular and organ function and dysfunction using direct and derived MRI methodologies}, pages 75--106, 2016.

\bibitem[Liu et~al.(2019)Liu, He, Zhang, and Liao]{liu2019statistical}
Xinlong Liu, Chu He, Qingyi Zhang, and Mingsheng Liao.
\newblock Statistical convolutional neural network for land-cover classification from sar images.
\newblock \emph{IEEE Geoscience and Remote Sensing Letters}, 17\penalty0 (9):\penalty0 1548--1552, 2019.

\bibitem[Presberger et~al.(2024)Presberger, Keshara, Stein, Kim, Grapin-Botton, and Andres]{presberger2024correlation}
Jannik Presberger, Rashmiparvathi Keshara, David Stein, Yung~Hae Kim, Anne Grapin-Botton, and Bjoern Andres.
\newblock Correlation clustering of organoid images.
\newblock \emph{arXiv preprint arXiv:2403.13376}, 2024.

\bibitem[Chopra et~al.(2005)Chopra, Hadsell, and LeCun]{chopra2005learning}
Sumit Chopra, Raia Hadsell, and Yann LeCun.
\newblock Learning a similarity metric discriminatively, with application to face verification.
\newblock In \emph{2005 IEEE computer society conference on computer vision and pattern recognition (CVPR'05)}, volume~1, pages 539--546. IEEE, 2005.

\bibitem[Goldberger et~al.(2004)Goldberger, Hinton, Roweis, and Salakhutdinov]{goldberger2004neighbourhood}
Jacob Goldberger, Geoffrey~E Hinton, Sam Roweis, and Russ~R Salakhutdinov.
\newblock Neighbourhood components analysis.
\newblock \emph{Advances in neural information processing systems}, 17, 2004.

\bibitem[Bromley et~al.(1993)Bromley, Guyon, LeCun, S{\"a}ckinger, and Shah]{bromley1993signature}
Jane Bromley, Isabelle Guyon, Yann LeCun, Eduard S{\"a}ckinger, and Roopak Shah.
\newblock Signature verification using a" siamese" time delay neural network.
\newblock \emph{Advances in neural information processing systems}, 6, 1993.

\bibitem[Balestriero and LeCun(2022)]{balestriero2022contrastive}
Randall Balestriero and Yann LeCun.
\newblock Contrastive and non-contrastive self-supervised learning recover global and local spectral embedding methods.
\newblock \emph{Advances in Neural Information Processing Systems}, 35:\penalty0 26671--26685, 2022.

\bibitem[Hadsell et~al.(2006)Hadsell, Chopra, and LeCun]{hadsell2006dimensionality}
Raia Hadsell, Sumit Chopra, and Yann LeCun.
\newblock Dimensionality reduction by learning an invariant mapping.
\newblock In \emph{2006 IEEE computer society conference on computer vision and pattern recognition (CVPR'06)}, volume~2, pages 1735--1742. IEEE, 2006.

\bibitem[Johnson et~al.(1990)Johnson, Moore, and Ylvisaker]{johnson1990minimax}
Mark~E Johnson, Leslie~M Moore, and Donald Ylvisaker.
\newblock Minimax and maximin distance designs.
\newblock \emph{Journal of statistical planning and inference}, 26\penalty0 (2):\penalty0 131--148, 1990.

\bibitem[Mak and Joseph(2018)]{mak2018support}
Simon Mak and V~Roshan Joseph.
\newblock Support points.
\newblock 2018.

\bibitem[He et~al.(2016)He, Zhang, Ren, and Sun]{he2016deep}
Kaiming He, Xiangyu Zhang, Shaoqing Ren, and Jian Sun.
\newblock Deep residual learning for image recognition.
\newblock In \emph{Proceedings of the IEEE conference on computer vision and pattern recognition}, pages 770--778, 2016.

\bibitem[Abadi et~al.(2016)Abadi, Barham, Chen, Chen, Davis, Dean, Devin, Ghemawat, Irving, Isard, et~al.]{abadi2016tensorflow}
Mart{\'\i}n Abadi, Paul Barham, Jianmin Chen, Zhifeng Chen, Andy Davis, Jeffrey Dean, Matthieu Devin, Sanjay Ghemawat, Geoffrey Irving, Michael Isard, et~al.
\newblock Tensorflow: A system for large-scale machine learning.
\newblock In \emph{12th $\{$USENIX$\}$ symposium on operating systems design and implementation ($\{$OSDI$\}$ 16)}, pages 265--283, 2016.

\end{thebibliography}

\end{document}